\documentclass[12pt]{article}

\usepackage{amsmath}
\usepackage{amssymb}
\usepackage{graphicx}
\usepackage{cite}
\usepackage{hyperref}

\begin{document}
\title{Model-independent features of gravitational waves from bubble collisions}
\author{%
Ariel M\'{e}gevand\thanks{Member of CONICET, Argentina. E-mail address:
megevand@mdp.edu.ar}~ 
and Federico Agust\'{\i}n Membiela\thanks{Member of CONICET, Argentina. E-mail
address: membiela@mdp.edu.ar} \\[0.5cm]
\normalsize \it IFIMAR (CONICET-UNMdP)\\
\normalsize \it Departamento de F\'{\i}sica, Facultad de Ciencias Exactas
y Naturales, \\
\normalsize \it UNMdP, De\'{a}n Funes 3350, (7600) Mar del Plata, Argentina }
\date{}
\maketitle
\begin{abstract}
We  study the gravitational radiation 
produced by the collisions of
bubble walls or thin fluid shells in cosmological phase transitions.
Using the so-called envelope approximation,
we obtain analytically the asymptotic behavior of the gravitational wave spectrum 
at low and high frequencies for any phase transition model. 
The complete
spectrum can thus be approximated by a simple interpolation between
these asymptotes. 
We verify this approximation with specific examples.
We use these results to discuss the dependence of the spectrum
on the time and size scales of the source. 
\end{abstract}

\section{Introduction}

In a phase transition of the Universe,
the disturbance produced in the hot plasma 
is a source of interesting phenomena such as 
baryogenesis \cite{krs85,ckn93} or 
the formation of  gravitational waves (GWs) \cite{tw90}. In particular, a phase
transition at the TeV scale gives naturally a GW spectrum that may
be observable by the space-based interferometer LISA \cite{LISA}. This fact has
motivated the investigation of GW production in the electroweak phase
transition, which may be strong enough in several extensions of the
Standard Model \cite{amnr02,m08,eknq08,hk08b,ak09,lms12,dhn14,kkm15,hkkm16,cns16,hknr16,lm16b,hwwcz16,gq16,ky16,hlw16,hkkkm17,alw17,v17,dhkn17,blwww17,mrv17,csw17,klmy17,kkm18,mnq18,aggsv19,crs19,mooy19,prs19,mmv19,ammpsv20,mp20,elpry20,bdfkm20}.
Gravitational waves generated in other phase transitions have also
been studied,
as well as their detectability prospects \cite{gs07,s15,dm16,jks16,jt16,bfmw17,b17,tyy17,csw18,hjkktt19,ms19,cgw19,bg19,bhk19,fhw19,abr20,eln20,abs20,f21,bj21}.
In general, a cosmological phase transition can be modeled with a scalar
order-parameter field $\phi(\mathbf{x},t)$ which couples to a 
plasma composed of several species of relativistic particles. In the
electroweak phase transition, this classical field represents the
expectation value of the Higgs field. 
The value $ \phi =0 $ corresponds to the symmetric, metastable phase, while 
a nonvanishing value corresponds to the stable, broken-symmetry phase.

In the case of a first-order phase transition, bubbles of the stable phase nucleate and expand
into the supercooled metastable phase.
A bubble is essentially a configuration in which the scalar field takes the stable-phase value  in a certain region
and vanishes outside.
The expansion of bubbles is driven by the pressure difference between the two phases. 
In most cases the bubble walls reach a terminal
velocity due to the friction with the plasma \cite{lmt92,t92,dlhll92,k92,a93,mp95l,mp95,mt97,js01,ms10,hs12,m13,knr14,k15,whz20,lc20}
and to hydrodynamic obstruction \cite{gkkm84,k85,kk86,eikr92,ikkl94,kl95,kl96,ms09,ekns10,kn11,lm16}.
However, there are scenarios in which the wall undergoes a continuous
acceleration or runaway behavior \cite{bm09,bm17,av21}, especially when
there is significant supercooling (see, e.g., \cite{mr17,eln19,elnv19,gwlisa2,bkmos19,whz20b}).
In any case, the variation of temperature due to the adiabatic cooling
or to reheating generally causes variations of the nucleation rate $\Gamma(t)$
and the wall velocity $v(t)$ as functions of time $t$ (see, e.g.,
\cite{ms08}).

A few different processes can produce GWs in a phase transition.
The bubble collision mechanism
is directly related to the propagation of the  bubble walls  \cite{ktw92b}. 
On the other hand, the walls cause
bulk fluid motions which may lead to gravitational radiation via 
turbulence \cite{kkt94,dgn02,kmk02,gkk07,cds09,kks10,kk15,nss18,pmbkk20}
or sound waves \cite{hhrw14,gm14,hhrw15,hhrw17,h18,hh19,gsvw21}, 
(see \cite{gwlisa}
for a review of these mechanisms). If the wall reaches a terminal
velocity, most of the energy released in the transition will go to
reheating and bulk fluid motions (see, e.g., \cite{ekns10,lm11}).
In such scenarios the GW signal is dominated by the fluid mechanisms.
On the other hand, in cases of continuous wall acceleration, 
an important fraction of the energy accumulates in the bubble walls
(see, e.g., \cite{ekns10,lm16}) and 
the bubble collisions become important.

The envelope approximation for the bubble collision mechanism
consists in modeling the bubble walls as infinitely-thin spherical
surfaces and considering only the uncollided  parts of them
as sources of GWs. 
The original calculation \cite{kt93} was based on a simulation in which
bubbles were nucleated at arbitrary points in space
and with a distribution in time corresponding to a nucleation rate
$\Gamma(t)\propto e^{\beta t}$, and their 
radii grew with a constant velocity $v$. 
This numerical computation was repeated in Refs.~\cite{hk08}
and \cite{w16} with technical improvements such as considering more
bubbles in the simulation. The resulting GW power spectrum has the
form of a broken power law in frequency. Specifically, the spectrum
rises as a power $\omega^{a}$ for low frequencies and falls as $\omega^{-b}$
for high frequencies, where $a$ is close to 3 and $b$ is close to
1. The peak frequency is of the order of the time scale $\beta^{-1}$.
Lattice simulations for the evolution of the scalar field 
have also been used to compute the GW spectrum from bubble collisions
\cite{cg12,w16,chw18,cghw20}.
The precise value of the peak of the spectrum is found to be slightly
shifted to lower frequencies with respect to the envelope approximation,
and the exponent of the high-frequency power law varies from $b\simeq1.5$
to $b\simeq2.3$, depending on the wall width.

The envelope approximation has also been used to compute the gravitational
radiation from bulk fluid motions, assuming that the fluid is concentrated
in thin shells next to the walls \cite{kkt94,hk08}. In Ref.~\cite{w16}, such a computation was compared
with a lattice simulation of the coupled system of scalar field and
fluid. It was shown that, for GWs generated by the fluid during bubble
collisions, the form of the spectrum is different for thick 
walls\footnote{Moreover, after bubble collisions, the acoustic and turbulent behaviors
of the fluid cannot be modeled by the envelope approximation \cite{h18}.}.
A more recent semi-analytic calculation \cite{jt17} 
for an exponentially growing
nucleation rate and a constant wall velocity
confirmed the broken
power law, with $a=3$ and $b=1$. 
In this approach, only two integrals must
be computed numerically, thus allowing to reach a wider frequency
range.  
A modification of the envelope approximation, the so-called
bulk flow model, consists in considering thin fluid shells which persist after the walls
collide. This model was investigated either with semi-analytical calculations 
\cite{jt19} and by simulating the formation and expansion of the
thin fluid shells  \cite{k18}.
Recently, we discussed a more general semi-analytic approach \cite{mm21a},
which can be applied to the envelope or bulk-flow approximations,
as well as to more general wall kinematics.

The relative simplicity of the envelope approximation is useful to
study the dependence of the GW spectrum on the phase transition model.
In the simulations of Ref.~\cite{w16}, a simultaneous nucleation
as well as an exponentially growing nucleation rate were considered.
In Ref.~\cite{jlst17}, the semi-analytical method of \cite{jt17}
was applied to a nucleation model of the form $e^{\beta t-\gamma^{2}t^{2}}$.
In this case, the exponential and simultaneous nucleations are obtained in the limits
of very low and very high $\gamma$, respectively. On the other hand, in the
lattice simulations of Ref.~\cite{chw18}, 
a constant nucleation rate was considered as well as
the exponential and simultaneous cases. 
The different spectra obtained in these works are qualitatively
similar, suggesting that the power laws at low and high frequencies
do not depend on the nucleation rate. This also seems to indicate
that the GW signal does not have a strong dependence on the distribution
of bubble sizes, which is quite different for different nucleation
rates.

It is worth mentioning that, for such a comparison between nucleation
rates, 
the energy
of the gravitational radiation is usually divided by the released
vacuum energy, and the frequency $\omega$ is divided by some characteristic
parameter $\omega_{*}$ which has the same meaning in the different
scenarios. For instance, using the average final bubble separation $d_{b}$
as a unit of frequency, $\omega_{*}=d_{b}^{-1}$ (see, e.g., \cite{chw18}),
the models under comparison have the
same value of $d_{b}$. The parameter $\beta$ of the exponential
rate can also be used as a unit. Although this quantity is rather artificial
for other models, it can be defined, e.g., by inverting
the relation which holds for the exponential
case, $d_{b}=(8\pi)^{1/3}v/\beta$ (this was used, e.g., in Ref.~\cite{gwlisa} to put the results
of Ref.~\cite{hhrw15} in terms of $\beta$). For comparing only
the shape of the spectrum, the peak frequency $\omega_{p}$ can be
used \cite{jlst17}. The precise choice of $\omega_{*}$ 
in terms of some length or time  associated to the phase transition kinematics
will determine the relative position of the peak between different models.

In the present paper we
use the envelope approximation to investigate
the dependence of the GW spectrum on specific features of the phase
transition, such as the nucleation rate and the wall velocity, and,
more generally, on length and time scales of the source. The bubble
collision mechanism is particularly suitable for that aim since it
is the one that links more directly the 
kinematics of bubble nucleation and expansion
to the GW spectrum. 
We also discuss a technique for finding the asymptotic behavior of the spectrum at high frequency.
We obtain analytically the
power laws $\omega^{3}$ and $\omega^{-1}$ for the envelope approximation
independently of $\Gamma(t)$ and $v(t)$. For the case of a constant
wall velocity, we obtain analytically the dependence on the parameter $ v $.

The plan of the paper is the following. 
In the next section we review   the development of a first-order phase transition
and we discuss the general definition of a characteristic
time scale for general forms of $\Gamma(t)$ and $v(t)$. 
In Sec.~\ref{bbcoll} we 
discuss the definition of a dimensionless GW spectrum which is suitable for model comparison
and we write down the expressions we shall use for the envelope approximation.
In Sec.~\ref{asymptotic} we investigate the form of
the spectrum at low and high frequencies. 
In Sec.~\ref{specific} we consider several specific cases, corresponding to a constant wall velocity  
and different nucleation rates
(namely, an exponential, a delta function, a Gaussian, and a constant rate). 
In Sec.~\ref{scales} we use the results to discuss the dependence of the
GW spectrum on the characteristics of the phase transition. 
We conclude with a discussion on the bubble collision mechanism in Sec.~\ref{conclu}. More details on the
calculations and on the numerical results, as well as analytic formulas
and comparisons with previous approaches are given in the appendices.

\section{General parametrization of bubble kinematics}

\label{gwwalls}

In the envelope approximation, one considers bubble walls which are spherical surfaces
(as bubbles overlap, the walls are assumed to disappear in the overlapping regions). 
In this picture, there is a homogeneous wall velocity $ v(t) $.
Thus, for a bubble nucleated at a certain time $t_{N}$, the radius
at time $t$ is given by 
\begin{equation}
	R(t_{N},t)=\int_{t_{N}}^{t}v(t'')dt'',\label{Rgral}
\end{equation}
where we have ignored for simplicity
the scale factor (which is a good approximation if the transition is short enough),
and we have assumed that the initial bubble size can be neglected (which is often the case). 
Assuming as well a homogeneous 
nucleation rate $\Gamma(t)$ per unit time per unit volume, and
taking into account bubble
overlapping, the average fraction of volume remaining in the high-temperature
phase at time $t$ is given by $f_{+}(t)=e^{-I(t)}$, with \cite{gt80,gw81,tww92}
\begin{equation}
I(t)=\int_{-\infty}^{t}dt''\Gamma(t'')\frac{4\pi}{3}R(t'',t)^{3}.\label{It}
\end{equation}
The nucleation rate actually  vanishes for $t<t_{c}$, where $t_{c}$ is the
time corresponding to the critical temperature, so the lower
limit of integration in Eq.~(\ref{It}) can be replaced by $t_{c}$.
However, doing so is somewhat misleading, since in most cases 
$ \Gamma(t) $ is actually negligible still at later times $t>t_{c}$,
so the quantity $I(t)$ does not really depend on the value of $t_{c}$.

The general form of the nucleation rate as a function of the temperature $ T $
is 
\begin{equation}
	\Gamma= A\exp[-S(T)],
\end{equation}
where $S$ is the instanton action. For
a vacuum transition \cite{c77,cc77}, $S$ is a constant and the factor
$A$ is of order $M^{4}$, where $M$ is the energy scale of the model.
For a thermal transition, we have $ A\sim T^4 $. In this case 
\cite{l81,l83}, $S$ has a strong dependence on the temperature,
the dynamics of nucleation is dominated by the exponential, and the specific form
of the prefactor is not too relevant. The adiabatic cooling of the Universe
causes in principle a rapid growth of $\Gamma$ with time. However,
depending on the global dynamics of the phase transition, $\Gamma$ may begin to
decrease at a certain point, as quickly as it previously grew. Two
possible scenarios for such a decrease are the system getting stuck
in the false vacuum (in the case of a very strong phase transition),
or a reheating of the plasma, which occurs when the phase transition
is mediated by slow deflagration bubbles (see Refs.~\cite{mr17,mr18}
for recent discussions).

In practice, bubble nucleation becomes noticeable at a certain time
$t_{*}$, after $\Gamma$ becomes of order $H^{4}$, where $H$
is the Hubble rate. Then, in general, bubbles fill all the space in a
short time $t_{b}\ll H^{-1}$. The kinematics of bubble nucleation
and growth may involve different characteristic times. For instance,
$\Gamma(t)$ may turn off in a relatively short time due to reheating,
after which bubble expansion may continue for a longer time  \cite{mr18}. 
We shall denote by $ t_\Gamma $ the time associated to bubble nucleation.
Without loss of generality, we can always define a dimensionless function $f(\tau)$ 
such that we can write
\begin{equation}
	\Gamma(t)=\Gamma_{*}\,f\left(\frac{t-t_{*}}{t_\Gamma}\right),\quad\mathrm{with}\quad f(0)=1,
	\label{gammaini}
\end{equation}
so that $\Gamma_{*}=\Gamma(t_{*})$ for a certain reference time $ t_* $.
Since in general $\Gamma(t)$
has a very rapid variation, the prefactor $ \Gamma_* $
is rather meaningless unless the time $t_{*}$ is inside, or very close
to, the time interval in which the phase transition
effectively occurs (i.e., where most bubbles nucleate and the fraction of volume $ f_+ $ has a significant variation).
The  number density of  bubbles, 
\begin{equation}
	n_b=\int_{-\infty}^{+\infty} \Gamma(t)f_+(t) dt, \label{nb}
\end{equation} 
defines a characteristic length scale $d_{b}\equiv n_{b}^{-1/3}$, which is  an estimate of the average distance between nucleation centers. 
For cases in which the nucleation rate reaches a maximum at a time $ t_m $ within the relevant time  interval
(we consider specific examples below),
a convenient choice for the parameter $ t_* $ is $ t_* =t_m$.
If  $ \Gamma $ does not have a maximum but grows indefinitely, 
the time $ t_* $ can be associated, e.g., to the maximum of the effective nucleation rate $ \Gamma (t)e^{-I(t)}$.
In any case, by definition of $ t_\Gamma $ we have $ n_b \sim \Gamma_* t_\Gamma$,
and we can write\footnote{We could actually define the parameters $ t_* $ and 
	$ t_\Gamma $ such that we have exactly $ \Gamma_* t_\Gamma =n_b $, so that
	we would just have $g(\tau)=f(\tau)$ in Eq.~(\ref{Gammabeta}). However, we want to have the freedom to choose the parameters conveniently
	for the simplicity of the expression for $ \Gamma(t) $. Therefore, we relax the condition $ f(0)=1 $ to $ g(0)\sim 1 $.}
\begin{equation}
	\Gamma(t)=\frac{1}{t_\Gamma d_b^{3}}\, g\left(\frac{t-t_{*}}{t_\Gamma}\right),
	\quad\mathrm{with}\quad g(0)\sim1
	\label{Gammabeta}
\end{equation}

The relation between the time parameter $ t_\Gamma $ and the distance parameter $ d_b $
depends on the global dynamics of the phase transition. In particular,  these parameters
may not be directly related through the velocity of bubble expansion.
As already mentioned, the duration of the phase transition $ t_b $ may differ from the nucleation time $ t_\Gamma $.
The time $ t_b $ is more directly related to the average bubble size $ d_b $ through the average bubble wall velocity.
We thus define a velocity parameter $v_{b}=d_{b}/t_{b}$. 
If the two time scales are different, it is convenient to define the parameter $ \alpha=t_b/t_\Gamma $ and the function
$ \tilde\Gamma(\tau)= \alpha g(\alpha\tau) $. Thus, we may write
Eq.~(\ref{Gammabeta})  in terms of $ t_b $ and $ v_b $, 
\begin{equation}
\Gamma(t)=\frac{1}{v_{b}^{3}t_{b}^{4}}\tilde{\Gamma}\left(\frac{t-t_{*}}{t_{b}}\right),
\quad\mathrm{with}\quad\tilde{\Gamma}(0)\sim t_b/t_\Gamma . \label{gammaadim}
\end{equation}
In the simplest cases, we have a single time scale, 
$ t_b\sim t_\Gamma $, so $ \tilde \Gamma(0)\sim 1 $. 
The different  parametrizations we have discussed are  useful for different purposes, and in the rest of this paper we shall use the form (\ref{gammaadim}). 

Let us consider a few simple examples which span the different possibilities for the relation between $ t_b $ and $ t_\Gamma $.

\subsection{Constant nucleation rate}

As mentioned above, for a vacuum phase transition the nucleation rate is a constant. 
In a physical particle-physics model, this scenario could arise in the case of extreme supercooling, i.e.,
if the system is stuck in the metastable phase even when the temperature is much smaller than the critical temperature.
However, in such a case the energy density is dominated by vacuum energy and the Universe undergoes inflation
(see, e.g., \cite{mr17}).
Hence, the dynamics of the phase transition departs from the more common scenario we wish to discuss here. 
For a thermal phase transition, a constant nucleation rate will hardly be a good approximation since the instanton action $ S(T) $ is very sensitive to temperature variations.
A scenario in which the temperature remains approximately constant arises
when bubbles expand as slow deflagrations, where the temperature outside the bubbles is heated up by shock fronts which carry away the released latent heat
(see, e.g.~\cite{mr18}).
In this case there is a reheated stage in which the temperature is approximately constant and homogeneous. 
However, this temperature is higher than in the previous pre-reheating stage, 
so this constant rate is vanishingly small in comparison. 
Hence, the bubble nucleation effectively occurs in a small time interval at the beginning of bubble expansion,
and a better approximation for $ \Gamma(t) $ is a Gaussian or a delta function.
In spite of this, the approximation of a constant nucleation rate is often
used in time-consuming computations such as lattice simulations, so we shall discuss it here.

In the parametrization (\ref{gammaadim}), this case corresponds to the limit of $ t_\Gamma\gg t_b $,
while the opposite case $ t_\Gamma\ll t_b $ corresponds to a delta-function rate (considered below).
This model requires also assuming  that the bubble nucleation turns on
at a certain time $t_{0}$.
Thus, we have $\Gamma=\Gamma_{0}\Theta(t-t_{0})$.
For a constant velocity $v$, a trivial calculation gives $I(t)=\frac{\pi}{3}v^{3}\Gamma_{0}(t-t_{0})^{4}$,
so the fraction of volume in the old phase is given by $f_{+}=e^{-\left[(t-t_{0})/t_{b}\right]^{4}}$,
with $t_{b}=(\frac{\pi}{3}v^{3}\Gamma_{0})^{-1/4}$. The parameter
$t_{b}$ is associated to the duration of the phase transition, and we may use a parametrization of the form
(\ref{gammaadim}), 
\begin{equation}
\Gamma(t)=\frac{1}{v^{3}t_{b}^{4}}\,\frac{3}{\pi}\Theta\left(\frac{t-t_{0}}{t_{b}}\right),
\label{Gammaconst}
\end{equation}
with $t_{*}=t_{0}$ and $\tilde{\Gamma}(\tau)=\frac{3}{\pi}\Theta(\tau)$.
The parameter $ d_b $ defined from the bubble number density is not exactly given by $ v t_b $. 
A simple calculation gives $ n_b= \int_{-\infty}^{+\infty}\Gamma(t)e^{-I(t)}dt=(3/\pi)\Gamma(5/4) v^{-3}t_b^{-3}$ (where the last $ \Gamma $ symbol represents the Euler gamma function).
Therefore, we have $ d_b\simeq 0.98 vt_b $ (i.e., the velocity parameter defined by $ v_b=d_b/t_b $ does not coincide exactly with the velocity $ v $).

\subsection{Exponential nucleation rate}

The exponential nucleation rate $\Gamma(t)=\Gamma_{*}e^{\beta(t-t_{*})}$
is obtained by linearizing the instanton action $S(T(t))$ at the time
$t_{*}$. For a constant velocity, this rate gives $I(t)=8\pi v^{3}\Gamma(t)/\beta^{4}$,
and the fraction of volume varies from the asymptotic value $f_{+}=1$
for $t\to-\infty$ to $f_{+}=0$ for $t\to\infty$. Nevertheless,
most of the variation occurs in a  time interval of order $\beta^{-1}$.
If $t_{*}$ is not close enough to this interval, then the parameter
$\Gamma_{*}$ will not give even the order of magnitude of $\Gamma(t)$
at the relevant times. 
Whatever the values of the original parameters $t_{*}$
and $\Gamma_{*}$, 
we may write $\Gamma(t)=\Gamma_{*}'e^{\beta(t-t_{*}')}$,
where the new and old parameters are related by  $\Gamma_{*}'=\Gamma_{*}e^{\beta(t_{*}'-t_{*})}$. A convenient
choice for $t_*'$ is the time $t_{e}$ for which $I(t_{e})=1$, i.e.,
when $f_{+}$ has decreased to $e^{-1}$. Indeed, at $t=t_{e}$ the
average nucleation rate $\Gamma(t)f_{+}(t)$, as well as the total uncollided
wall area $\langle S_{\mathrm{tot}}(t)\rangle$, take their maximum
\cite{mm20}. By definition of $ t_e $ we have 
$I(t)=e^{\beta(t-t_{e})}$, so we may write 
\begin{equation}
\Gamma(t)=\frac{\beta^{4}}{8\pi v^{3}}e^{\beta(t-t_{e})}.
\label{nuclexpadim}
\end{equation}
Taking into account the well known relation $ d_{b}=(8\pi)^{1/3}v/\beta $,
we have
$\Gamma(t)=\beta d_b^{-3}e^{\beta(t-t_{e})}$, 
which is of the form (\ref{Gammabeta}) with $ t_\Gamma=\beta ^{-1}$.
If we define $t_{b}=t_\Gamma=\beta^{-1}$, 
Eq.~(\ref{nuclexpadim}) is also of the form (\ref{gammaadim}) 
and we have $\tilde{\Gamma}(\tau)=e^{\tau}/8\pi$.
Here, we have $ \tilde \Gamma(0) =1/8\pi\neq t_b/t_\Gamma$, since a different time parameter $t_{b}'>\beta^{-1}$
would actually be more representative of the duration of the phase transition
(see, e.g., \cite{mm20}). 
Nevertheless, we shall use  the form (\ref{nuclexpadim}) since $ \beta $ is the standard parameter.

\subsection{Gaussian nucleation rate}

As already mentioned, there are at least two different scenarios in which the nucleation rate may reach a maximum and
turn off during the phase transition:
\begin{itemize}
	\item[A.] Strong supercooling: $ S(T) $ has a minimum. 
	\item[B.] Reheating: $ T(t) $ has a minimum. 
\end{itemize}
Case A occurs when
a barrier between the minima of the effective potential persists at low temperatures \cite{mr17}.
In such a case, the nucleation rate initially grows as the temperature descends from the critical temperature $ T_c $
and the minima become non-degenerate. 
However, at low enough temperature 
the barrier between phases cannot be surpassed and the nucleation rate begins to decrease with decreasing temperature. 
Correspondingly, the instanton action $ S(T) $ has a minimum at a certain temperature $ T_m $.
Since $ T $ decreases as a function of time, this minimum will be reached at a certain time $ t_m $ (unless the phase transition is completed before that time). 
Expanding $ S(t) $ around its minimum, we obtain a Gaussian approximation for the nucleation rate,
\begin{equation}
	\Gamma(t)=\Gamma_{m}\exp[-\gamma^{2}(t-t_{m})^{2}].
	\label{gammagaussini}
\end{equation}
Case B occurs when a phase transition is mediated by slow deflagrations \cite{mr18}.
In the general scenario 
there is little supercooling, since the barrier between minima disappears at a temperature which is close to the critical one, and
in this range $ S(T) $ is a monotonous function.
However, for walls which propagate as deflagrations, the plasma outside the bubbles is reheated during the phase transition.
As a result, the temperature initially decreases due to the adiabatic expansion of the Universe, but at some point it begins to increase due to reheating. 
As a consequence, the temperature $ T(t) $ has a minimum at a certain time $ t_m $, and so does the
function $ S(T(t)) $,  so  the nucleation rate can be approximated again by Eq.~(\ref{gammagaussini}).

In case B, the maximum of the nucleation rate is always reached during the phase transition,
since the very existence of a minimum of $ T(t) $ is due to the reheating during bubble expansion.
In contrast, in case A the function $ \Gamma(T) $ has a maximum at a temperature $ T_m $ which may not be reached during the phase transition.
This will happen  if $ \Gamma_m $ is very large compared to $ H^4 $.
In such a case, the phase transition will complete at an earlier 
time $ t_* $ such that $ \Gamma(t_*) \sim H(t_*)^4$.
If this is the case, it is not a good approximation to expand $ S(T) $ at $ T_m $.  
Expanding at a higher temperature $ T_* $ will give a linear term, while the quadratic term is a second order correction.
Hence an exponential nucleation rate will not be a bad approximation.
This case was considered in Ref.~\cite{jlst17}, and we discuss it in some detail in App.~\ref{apgauss}.
On the other hand, in cases for which the maximum of $\Gamma(t)$ is reached during the phase 
transition\footnote{It is worth commenting that, in case A, if $ \Gamma_m $ is too low in comparison with the expansion parameter $ H^4 $, the phase transition will never complete 
	(see \cite{mr17} for details).},
we have a ``true Gaussian rate'', i.e., it cannot be approximated by an exponential rate.

The nucleation rate (\ref{gammagaussini}) is of the form (\ref{gammaini}), with $ t_* =t_m $, $ \Gamma_*=\Gamma_m $, and $ t_\Gamma = \gamma^{-1} $.
An interesting difference from the previous cases is that, since the nucleation rate turns off, there is a bound on the number of nucleated bubbles,
namely,
\begin{equation}
	n_{\max}=\int_{-\infty}^{+\infty} \Gamma(t)dt =\sqrt{\pi}\Gamma_{m}/\gamma.
\end{equation}
The actual number density (\ref{nb})  contains a factor $ f_+(t) $, which implies $ n_b < n_{\max} $. 
This bound defines a minimal bubble separation, $ d_{\min} =n_{\max}^{-1/3} $.
Unless the phase transition finishes before the maximum of the Gaussian is reached,  the value $ n_{\max} $ will be a good approximation for $ n_b $, and we have $ d_b\simeq d_{\min} $ (see App.~\ref{apgauss} for more details).
In terms of this parameter, Eq.~(\ref{gammagaussini}) becomes
\begin{equation}
	\Gamma(t)=(\gamma/\sqrt{\pi}d_{\min}^{3}) \exp[-\gamma^{2}(t-t_{m})^{2}],
	\label{gammagaussalpha}
\end{equation}
which is of the form (\ref{Gammabeta}) with $g(\tau)= e^{-\tau^2}/\sqrt{\pi}$.
The time $ t_b=d_b/v_b $ may be different from the nucleation time $ t_\Gamma $
(in particular, the phase transition may go on after the nucleation rate turns off).
In order to write Eq.~(\ref{gammagaussalpha}) in the form (\ref{gammaadim}), 
we shall use the analytic parameter $ t_{\min} =d_{\min}/v_b$ instead of $ t_b $ which must be obtained numerically.
We have
\begin{equation}
	\Gamma(t)=\frac{1}{v_b^3 t_{\min}^{4}} \tilde\Gamma\left(\frac{t-t_{m}}{t_{\min}}\right) ,
	\label{gammagauss}
\end{equation}
where
$ \tilde \Gamma(\tau)=\gamma t_{\min} \,g(\gamma t_{\min}\, \tau) $.
Since we have two different time scales, the dimensionless nucleation rate depends on the parameter 
$ \alpha\equiv \gamma t_{\min}=t_{\min}/t_\Gamma $.

\subsection{Delta-function nucleation rate}

If the time during which nucleation occurs is much shorter than the
total duration of the phase transition, the
nucleation rate can be approximated by a delta function $\Gamma(t)=n_{b}\delta(t-t_{*})$,
where $n_{b}$ is the number density of bubbles. 
This can be regarded as a limit of the Gaussian rate, and is a good approximation for 
some models of type B (in the classification of the previous subsection). In particular, 
when a sudden reheating of the plasma causes the nucleation rate to quickly turn off \cite{mr18}. 
Since the nucleation in this case
is simultaneous, the fundamental parameter is the distance
scale $d_{b}\equiv n_{b}^{-1/3}$. Using the well-known scaling property
of the delta distribution, we may write 
\begin{equation}
\Gamma(t)=\frac{1}{t_{b}d_{b}^{3}}\,\delta\left(\frac{t-t_{*}}{t_{b}}\right)\label{nuclsimultadim}
\end{equation}
for any parameter $ t_b $. The convenient time parameter here is the 
typical time of bubble growth. 
Given a characteristic (average) velocity $ v_b $, we have $t_{b}=d_{b}/v_b$. 
Hence, Eq.~(\ref{nuclsimultadim}) is  of the form (\ref{gammaadim}) with $ \tilde\Gamma(\tau)=\delta(\tau) $.
This can also be obtained as the limit for $ \alpha\to\infty $ of the Gaussian case
$ \tilde \Gamma(\tau)=\alpha \,g(\alpha\, \tau) $.

\section{Gravitational waves}
\label{bbcoll}

The gravitational wave power spectrum is often represented by the quantity
\begin{equation}
	\Omega_{GW}=\frac{1}{\rho_{\mathrm{tot}}}\frac{d\rho_{GW}}{d\ln\omega},\label{OmGW}
\end{equation}
i.e., the energy density  in gravitational radiation per logarithmic
frequency, divided by 
the total energy density of the Universe, $\rho_{\mathrm{tot}}$.
Before proceeding to the calculation of this quantity, we shall discuss
the definition of a dimensionless quantity which is useful for expressing general results and for model comparison.

\subsection{Dimensionless GW spectrum}

The quantity (\ref{OmGW})
is sometimes written in the form (see, e.g., \cite{hk08,jt17})
\begin{equation}
\Omega_{GW}=\kappa^{2}\left(\frac{H}{\beta}\right)^{2}\left(\frac{\alpha_{T}}{\alpha_{T}+1}\right)^{2}\Delta(\omega/\beta),\label{OmGWDelta}
\end{equation}
where  $\beta$ is the parameter of the exponential
nucleation rate, $\alpha_{T}$ is the ratio of the energy released
at the phase transition to the radiation energy, $\alpha_{T}=\rho_{\mathrm{vac}}/\rho_{\mathrm{rad}}$, 
$\kappa$ is an efficiency factor  \cite{kkt94} quantifying the fraction of
the released energy which goes into the source of GWs, and
the dimensionless function $\Delta$ is defined as 
\begin{equation}
\Delta(\omega/\beta)\equiv\frac{3\beta^{2}}{8\pi G(\kappa\rho_{\mathrm{vac}})^{2}}\frac{d\rho_{GW}}{d\ln\omega}(\omega).\label{Deltadef-exp}
\end{equation}
In these expressions,
the quantities $\kappa^{2}$, $ \rho_{\mathrm{vac}}^2 $, and $\beta^{2}$ are introduced just by multiplying and dividing them in Eq.~(\ref{OmGW}).
The other quantities
are introduced
by using the relation $H^{2}=8\pi G\rho_{\mathrm{tot}}/3$, assuming
that the total energy density can be decomposed into vacuum and radiation
energy densities, $\rho_{\mathrm{tot}}=\rho_{\mathrm{vac}}+\rho_{\mathrm{rad}}$,
and assuming that the vacuum energy density coincides with the latent
heat released at the phase transition. These approximations can be
improved (see \cite{cgstw21,gsvw21b} for recent discussions), but are useful to focus on the calculation
of the dimensionless quantity $\Delta$ for a simplified phase transition kinematics
and then applying Eq.~(\ref{OmGWDelta})
to specific realistic models (see, e.g., \cite{gwlisa,gwlisa2}).

Under suitable approximations,
the quantity $(\kappa\rho_{\mathrm{vac}})^{2}$ is a constant which 
will appear explicitly
in the expression for $d\rho_{GW}/d\ln\omega$ and 
cancel out in Eq.~(\ref{Deltadef-exp}), as well as the numerical constants.
On the other hand, using the parameter $\beta$ makes sense only for
the exponential nucleation rate, since for other cases the expression
for $d\rho_{GW}/d\ln\omega$ will depend on a different quantity.
Nevertheless, we may generalize the definition of $\Delta$ in terms of a more general reference frequency 
$ \omega_* $, 
\begin{equation}
\Delta(\omega/\omega_{*})\equiv\frac{3\omega_{*}^{2}}{8\pi G(\kappa\rho_{\mathrm{vac}})^{2}}\frac{d\rho_{GW}}{d\ln\omega}(\omega).\label{Deltadef}
\end{equation}
For a given mechanism of GW generation, the parameter $ \omega_{*} $ can be conveniently associated to a relevant time or length 
scale\footnote{It is worth noticing that this characteristic frequency determines the peak of the spectrum at the time of GW generation. The frequency, as well as the energy density, are subject to resdshifting.}.
Thus, for bubble collisions,  it is convenient to use the frequency $\omega_{b}= t_{b}^{-1}$ associated to the 
time parameter which appears explicitly in the parametrization (\ref{gammaadim}) and depends on the specific phase transition model.
However, for comparing two different models a single frequency unit must be used.
The relation between the dimensionless spectrum for
two different reference frequencies  is 
$ \Delta_2(\omega/\omega_2)= (\omega_2/\omega_1)^2 \Delta_1 ((\omega_2/\omega_1)(\omega/\omega_2))$.

\subsection{GWs from bubble walls}

We shall use the approach of Ref.~\cite{mm21a}, which we summarize very briefly.
For a large volume $ V $, the GW power spectrum is  written in the form 
\begin{equation}
	\frac{d\rho_{GW}}{d\ln\omega}=\frac{4G\omega^{3}}{\pi}\int_{-\infty}^{\infty}dt\int_{t}^{\infty}dt'\cos[\omega(t-t')]\,\Pi(t,t',\omega),\label{rhogw}
\end{equation}
where
\begin{equation}
	\Pi(t,t',\omega)\equiv\frac{1}{V} \Lambda_{ij,kl}(\hat{n})\left\langle\tilde{T}_{ij}(t,\omega\hat{n})\tilde{T}_{kl}(t',\omega\hat{n})^{*}\right\rangle ,
	\label{defPi}
\end{equation}
$ \Lambda_{ij,kl} $ is the transverse-traceless projection tensor for the direction of observation $ \hat n $, 
$ \tilde T_{ij} $ is the spatial Fourier transform of the stress-energy tensor $ T_{ij} $ of the source, and $ \langle~\rangle $ indicates ensemble average.
If $ T_{ij} $ is decomposed as a sum over bubbles, $ \Pi $
naturally separates as $\Pi=\Pi^{(s)} +\Pi^{(d)}$, where $ \Pi^{(s)} $ contains correlations between different points on a single bubble
and $ \Pi^{(d)} $ contains correlations between two different bubbles (such a separation also arises in the treatment of Ref.~\cite{jt17}).
For gravitational waves from bubble walls, $ T_{ij} $ is approximated by a surface delta function which eliminates
some of the spatial integrals in the Fourier transforms $ \tilde T_{ij},\tilde T_{kl} $. 
In the case of the envelope approximation we have, for each bubble, 
\begin{equation}
	T_{ij}=\sigma\delta(r-R)\,\hat{r}_{i}\hat{r}_{j}\,1_{S}(\hat{r}),
	\label{Tijenv}
\end{equation}  
where $ \sigma $ is the surface energy density, $ r $ is the distance from the bubble center, $ R $ is the bubble radius, $ R\hat r $ is the position of a point on the bubble surface,
and $1_{S}$ is the indicator function for the uncollided wall. 
To take into account the energy which accumulates in the wall, the usual replacement
$\sigma=(\kappa\rho_{\mathrm{vac}}/3)R$ is made,
where the efficiency factor $\kappa$ accounts for the fraction
of energy which goes either to the wall (in a vacuum phase transition we have $ \kappa =1 $)
or to bulk fluid motions (which are assumed to occur in thin shells next to the walls)
(see, e.g., \cite{m08,ekns10,lm11,lm15,lm16,lm16b,elnv19,elv20,gkv20,gkv21} for the calculation of this factor). 
Finally, the sum over bubbles and the statistical average are related to the nucleation rate $ \Gamma(t) $, 
and several of the remaining angular integrals can be performed analytically.

The result depends on the probability that  two points  at angular positions 
$ \hat r,\hat r' $ on the bubble surfaces at times $ t $ and $ t' $ are both uncollided.
This probability was studied in Ref.~\cite{mm20}. It is proportional to 
$ e^{-I(t)} e^{-I(t')} e^{I_\cap}$, where the last factor 
takes into account the fact that the probabilities for the two points are not 
independent\footnote{If the points belong to the surfaces of two different bubbles,
	the probability includes also Heaviside functions
	which vanish if the bubbles are so close that one of the points has been captured by the other bubble.}. We have
\begin{equation}
	I_{\cap}(t,t',s)=\int_{-\infty}^{t}dt''\Gamma(t'')V_{\cap}(t'',t,t',s),\label{Itts}
\end{equation}
where $s$ is the distance between the points and 
\begin{equation}
	V_{\cap}=\frac{\pi}{12}(r+r'-s)^{2}\left[s+2(r+r')-\frac{3(r-r')^{2}}{s}\right]\Theta(r+r'-s).\label{VI}
\end{equation}
Here, $\Theta$ is the Heaviside step function, and we have used the notation $r=R(t'',t)$, $r'=R(t'',t')$
(for more details and interpretation, see \cite{mm20} or \cite{mm21a}).
The final expressions from Ref.~\cite{mm21a} (see \cite{jt19} for similar expressions) are
\begin{equation}
	\frac{\Pi^{(s)}(t,t',\omega)}{(\kappa\rho_{\mathrm{vac}}/3)^2}=\frac{\pi^{2}}{4}\int_{-\infty}^{t}dt_{N} \Gamma(t_{N})
	\int_{R_{-}}^{R_{+}}\frac{ds}{s^{3}}\, \sum_{i=0}^{2}P_{i}(R_{+},R_{-},s)\frac{j_{i}(\omega s)}{(\omega s)^{i}}e^{-I_{\mathrm{tot}}(t,t',s)},
	\label{Piscolfinal}
\end{equation}
\begin{align}
	\frac{\Pi^{(d)}(t,t',\omega)}{(\kappa\rho_{\mathrm{vac}}/3)^2}= & \frac{\pi^{3}}{4}\int_{-\infty}^{t}dt_{N}\Gamma(t_{N})
	\int_{-\infty}^{t'}dt_{N}^{\prime}\Gamma(t_{N}^{\prime})\nonumber \\
	& \times\int_{R_{-}}^{R_{+}-|R(t_{N},t_{N}')|}\frac{ds}{s^{4}}\,
	e^{-I_{\mathrm{tot}}(t,t',s)}\frac{j_{2}(\omega s)}{(\omega s)^{2}}Q_{+}(s,R,R_{-})Q_{-}(s,R',R_{-}),
	\label{Pidcolfinal}
\end{align}
where $ I_\mathrm{tot} = I(t)+I(t')-I_\cap(t,t',s)$,  
the $ j_i $ are spherical Bessel functions, 
\begin{equation}
	j_{0}(x)=\frac{\sin x}{x},\quad j_{1}(x)=\frac{\sin x-x\cos x}{x^{2}},\quad j_{2}(x)=\frac{(3-x^{2})\sin x-3x\cos x}{x^{3}},\label{SBessel}
\end{equation}
and the $P_{i}$  and $ Q_\pm $ are polynomials in $ R$ and $R' $, which have simpler expressions in terms of the variables
$R_{+}=R'+R$ and\footnote{Notice that $ R_-=\int_{t}^{t'}v_{w}(t'')dt'' $ is given by
	$R'-R$ only for 
	the single-bubble case (for the two-bubble case 
	the latter difference depends on the nucleation times $ t_N,t_N' $).}
$R_{-}=R(t,t')$,
\begin{align}
	P_{0}(R_{+},R_{-},s)= & \,(s^{2}-R_{-}^{2})^{2}(s^{2}-R_{+}^{2})^{2},\label{P0}\\
	P_{1}(R_{+},R_{-},s)= & \,2(s^{2}-R_{-}^{2})(s^{2}-R_{+}^{2})\left[3s^{4}+s^{2}(R_{-}^{2}+R_{+}^{2})-5R_{-}^{2}R_{+}^{2}\right],\\
	P_{2}(R_{+},R_{-},s)= & \,3s^{8}+2s^{6}(R_{-}^{2}+R_{+}^{2})+3s^{4}(R_{-}^{4}+4R_{-}^{2}R_{+}^{2}+R_{+}^{4})\label{P2}\\
	& -30s^{2}R_{-}^{2}R_{+}^{2}(R_{-}^{2}+R_{+}^{2})+35R_{-}^{4}R_{+}^{4}\nonumber 
\end{align}
and 
\begin{align}
	Q_{+} & (s,R,R_{-})=(s^{2}-R_{-}^{2})\left[(2R+R_{-})^{2}-s^{2}\right]\left[s^{2}-R_{-}(2R+R_{-})\right],\label{Qmas}\\
	Q_{-} & (s,R',R_{-})=(s^{2}-R_{-}^{2})\left[(2R'-R_{-})^{2}-s^{2}\right]\left[s^{2}+R_{-}(2R'-R_{-})\right].\label{Qmenos}
\end{align}

Now, we insert these results 
in the GW energy density (\ref{rhogw}) and then in 
the dimensionless spectrum (\ref{Deltadef}). 
We obtain $ \Delta = \Delta^{(s)} + \Delta^{(d)} $, with
\begin{align}
\Delta^{(s)}= & \,\frac{\omega^{3}\omega_{*}^2}{48}\int_{-\infty}^{\infty}dt_{+}\int_{0}^{\infty}dt_{-}\cos(\omega t_{-})\int_{-\infty}^{t}dt_{N}\Gamma(t_{N})\nonumber \\
 & \times\int_{R_{-}}^{R_{+}}\frac{ds}{s^{3}}e^{-I_{\mathrm{tot}}(t,t',s)}\sum_{i=0}^{2}\frac{j_{i}(\omega s)}{(\omega s)^{i}}\,P_{i}(R_{+},R_{-},s),\label{Deltaenvs}
\end{align}
\begin{align}
\Delta^{(d)} & =\frac{\pi\omega^{3}\omega_*^2}{48}\int_{-\infty}^{\infty}dt_{+}\int_{0}^{\infty}dt_{-}\cos(\omega t_{-})\int_{-\infty}^{t}dt_{N}\Gamma(t_{N})\int_{-\infty}^{t'}dt_{N}'\Gamma(t_{N}')\nonumber \\
 & \times\int_{R_{-}}^{R_{+}-|R(t_{N},t_{N}')|}\frac{ds}{s^{4}}e^{-I_{\mathrm{tot}}(t,t',s)}\frac{j_{2}(\omega s)}{(\omega s)^{2}}Q_{+}(s,R,R_{-})Q_{-}(s,R',R_{-}).\label{Deltaenvd}
\end{align}
where we have changed the variables $t,t'$ to $t_{\pm}=t'\pm t$.

\section{Asymptotic behavior}

\label{asymptotic}

Before considering specific examples, we shall study the general behavior
of the GW spectrum   for low and high frequencies.

\subsection{Low frequency}

A general argument based on causality shows that for a transient stochastic source 
the low frequency tail of the GW spectrum $ \Omega_{GW} $ is proportional to $ \omega^3 $
(see \cite{cdks09} and the more recent review \cite{cf18}). 
As already mentioned, this power law has been verified numerically for the bubble collision mechanism in the envelope approximation. 
It is worth mentioning that this needs not be the case for every mechanism of GW generation related to  the phase transition. 
As an example, for the bulk flow model with  long-lasting fluid shells
the GW spectrum for small $ \omega $ behaves as $ \propto \omega $. 
This was shown analytically in Ref.~\cite{jt19}, and we shall use similar considerations for the envelope approximation.
In this case, the source of GWs turns
off as soon as the phase transition ends and the bubble walls disappear.

In Eqs.~(\ref{Deltaenvs})-(\ref{Deltaenvd}), the time variables $t_{\pm}$ 
have an effective range of order $t_{b}$ around the time $t_{*}$,
since the exponential $e^{-I_{\mathrm{tot}}}$, like $ e^{-I(t)} $, becomes negligible at later times.
As a consequence, the spatial variable $s$ is bounded by $\sim  v_{b}t_{b}$. 
Hence, for $\omega\ll t_{b}^{-1}$,
the oscillating functions in the integrand can be expanded in powers
of $\omega$. The zeroth order corresponds to the replacements
\begin{equation}
\cos(\omega t_{-})\rightarrow1,\:j_{0}(\omega s)\rightarrow1,\:\frac{j_{1}(\omega s)}{\omega s}\rightarrow\frac{1}{3},\:\frac{j_{2}(\omega s)}{(\omega s)^{2}}\rightarrow\frac{1}{15},\label{depwLF}
\end{equation}
and the quantity $\Delta$ becomes, at low frequencies, 
\begin{equation}
\Delta_{LF}=\Delta_{LF}^{(s)}+\Delta_{LF}^{(d)}=\left(B^{(s)}+B^{(d)}\right)\omega^{3},\label{DeltaLFgral}
\end{equation}
where\footnote{We remark that the general definition of $R_{-}$ is $R_{-}=R(t,t')$,
and the expression $R_{-}=R'-R$ is valid only for the single-bubble
case.} 
\begin{equation}
B^{(s)}=\frac{\omega_{*}^{2}}{48}\int_{-\infty}^{\infty}dt_{+}\int_{0}^{\infty}dt_{-}\int_{-\infty}^{t}dt_{N}\Gamma(t_{N})\int_{R_{-}}^{R_{+}}\frac{ds}{s^{3}}\,\left[P_{0}+\frac{P_{1}}{3}+\frac{P_{2}}{15}\right]e^{-I_{\mathrm{tot}}},\label{Bgrals}
\end{equation}
 and 
\begin{equation}
B^{(d)}=\frac{\pi \omega_{*}^{2}}{48}\int_{-\infty}^{\infty}\negmedspace dt_{+}\int_{0}^{\infty}\negmedspace dt_{-}\int_{-\infty}^{t}\negmedspace dt_{N}\Gamma(t_{N})\int_{-\infty}^{t'}\negmedspace dt_{N}'\Gamma(t_{N}')\int_{R_{-}}^{R_{+}-|R(t_{N},t_{N}')|}\negmedspace ds\frac{Q_{+}Q_{-}}{15s^{4}}e^{-I_{\mathrm{tot}}}.\label{Bgrald}
\end{equation}

\subsection{High frequency}

For $\omega\gg t_{b}^{-1}$, all the quantities appearing in Eqs.~(\ref{Deltaenvs})-(\ref{Deltaenvd})
have a slow variation in comparison with the oscillating functions
$\cos(\omega t_{-})$ and $j_{i}(\omega s)$. We change to variables
$x=\omega s$ and $y=\omega t_{-}$ in order to eliminate the frequency
from the latter, and then we define $\epsilon\equiv1/\omega$
and expand the quantities in powers of $\epsilon$. In the first place,
we have 
\begin{equation}
t=t_{+}/2-\epsilon y/2,\quad t'=t_{+}/2+\epsilon y/2,\quad s=x\epsilon.\label{HFcv}
\end{equation}
 For the bubble radius (\ref{Rgral}), we obtain 
\begin{equation}
R=\bar{R}-\epsilon\,\bar{v}\,y/2+\mathcal{O}(\epsilon^{2}),\quad R'=\bar{R}'+\epsilon\,\bar{v}\,y/2+\mathcal{O}(\epsilon^{2}),
\end{equation}
where $\bar{R}\equiv R(t_{N},\bar t )$, $\bar{R}'\equiv R(t_{N}',\bar t)$, $\bar{v}\equiv v(\bar t)$, and $ \bar t=t_+/2 $
($ \bar R $ and $ \bar R' $ are equal for the single-bubble case).
Hence, we have\footnote{For any smooth function $f(t)$, we have $f(t')-f(t)=f'(\bar t)\epsilon y+\mathcal{O}(\epsilon^{3})$
and $f(t)+f(t')=2f(\bar t)+f''(\bar t)\epsilon^{2}y^{2}/4+\mathcal{O}(\epsilon^{4})$.
We use these identities a couple of times below.} $R_{-}=\epsilon\,\bar{v}\,y+\mathcal{O}(\epsilon^{3})$, $R_{+}=2\bar{R}+\mathcal{O}(\epsilon^{2})$.
From Eq.~(\ref{It}) we obtain
\begin{equation}
I(t)+I(t')=2I(\bar t)+O(\epsilon^{2})
\end{equation}
and, from Eqs.~(\ref{Itts}) and (\ref{VI}),
\begin{equation}
I_{\cap}(t,t',s)=I(\bar t)-\epsilon\pi\frac{x^{2} + \bar{v}^{2} y^{2}}{x} I_{2}(\bar t) + O(\epsilon^{2}),\label{IintHF}
\end{equation}
where we have used the notation 
\begin{equation}
I_{n}(t)=\int_{-\infty}^{t}dt''\Gamma(t'')R(t'',t)^{n} \label{In}
\end{equation}
(the function $I_{3}$ is proportional to $I$). We thus have
\begin{equation}
e^{-I_{\mathrm{tot}}}=e^{-I(\bar t)}\left[1-\epsilon\pi I_{2}(\bar t) \frac{x^{2}+\bar{v}^{2}y^{2}}{x}+\mathcal{O}(\epsilon^{2})\right].\label{eItotHF}
\end{equation}

Let us consider first the single-bubble contribution, 
\begin{align}
\Delta^{(s)} & = \frac{\epsilon^{-4}\omega_*^2}{24} \int_{-\infty}^{\infty}d\bar t\int_{0}^{\infty}dy\cos y 
\int_{-\infty}^{\bar t -\epsilon y/2}dt_{N}\Gamma(t_{N})\nonumber \\
 & \times\int_{\bar{v}y+\mathcal{O}(\epsilon^{2})}^{2\bar{R}/\epsilon+\mathcal{O}(\epsilon)}\frac{dx}{x^{3}}\sum_{i=0}^{2}\frac{j_{i}(x)}{x^{i}}\,P_{i}e^{-I(\bar t)}\left[1-\epsilon\pi I_{2}(\bar t)\frac{x^{2}+\bar{v}^{2}y^{2}}{x}+O(\epsilon^{2})\right]\label{DeltasenvHF}
\end{align}
The first term inside the brackets gives a vanishing contribution
upon integrating the variable $x$.\footnote{Indeed, we have $\int_{R_{-}}^{R_{+}}\frac{ds}{s^{3}}\sum_{i=0}^{2}\frac{j_{i}(\omega s)}{(\omega s)^{i}}\,P_{i}=0$
(to all order in $\epsilon$), which is a consequence of the fact that $\int d\hat{r}\int d\hat{r}'\,e^{i\omega\hat{n}\cdot\mathbf{s}}\Lambda_{ijkl}\hat{r}_{i}\hat{r}_{j}\hat{r}'_{k}\hat{r}'_{l}=0$.
The reason is that the approximation $I_{\mathrm{tot}}=I(\bar t)$
restores the spherical symmetry, since only the dependence of $I_{\cap}$
on the variable $s$ carries the information on the correlation between
different points on the walls.} Therefore, the bracket gives a factor of $\epsilon$. Besides, it
is easy to see that the polynomials $P_{i}$, Eqs.~(\ref{P0})-(\ref{P2}),
are of order $\epsilon^{4}$, 
\begin{equation}
P_{i}(R_{+},\epsilon y,\epsilon x)=16\,\epsilon^{4}\,\bar{R}^{4}\,p_{i}(x,\bar{v}y)\label{HFfs3}
\end{equation}
with 
\begin{equation}
p_{0}=(x^{2}-\bar{v}^{2}y^{2})^{2},\,p_{1}=-2(x^{2}-\bar{v}^{2}y^{2})(x^{2}-5\bar{v}^{2}y^{2}),\,p_{2}=3x^{4}-30x^{2}\bar{v}^{2}y^{2}+35\bar{v}^{4}y^{4}.
\end{equation}
Therefore, we have $\Delta^{(s)}\sim\epsilon=\omega^{-1}$. To 
this lowest order, we take the zeroth order in the limits of the integrals
in Eq.~(\ref{DeltasenvHF}). In this limit, the integration over
the nucleation time only affects the factor $\bar{R}^{4}$, and gives
a factor $I_{4}(\bar t)$. Interchanging the
order of the integrals with respect to $x$ and $y$, we obtain
\begin{align}
\Delta^{(s)}= & \,-\frac{2\pi}{3}\omega_{*}^{2}\omega^{-1}
\int_{-\infty}^{\infty}d\bar t e^{-I(\bar t)}I_{2}(\bar t)I_{4}(\bar t)\nonumber \\
 & \times\int_{0}^{\infty}\frac{dx}{x^{3}}\sum_{i=0}^{2}\frac{j_{i}(x)}{x^{i}}\int_{0}^{x/\bar{v}}dy\cos y\,(x^{2}+\bar{v}^{2}y^{2})\,p_{i}(x,\bar{v}y)+\mathcal{O}(\omega^{-2}).\label{DeltasenvHF-1}
\end{align}
The integrations on $x$ and $y$ can be done analytically, and we
obtain 
\begin{equation}
\Delta_{HF}^{(s)}=\omega^{-1}\,\frac{\pi \omega_{*}^{2}}{72}\int_{-\infty}^{\infty}d\bar t I_{4}(\bar t) I_{2}(\bar t) e^{-\frac{4\pi}{3}I_{3}(\bar t)}A^{(s)}(\bar{v})\label{HFAs}
\end{equation}
(where the notation $ HF $ indicates the high frequency limit), with 
\begin{equation}
A^{(s)}(\bar{v})=2\frac{3-11\bar{v}^{2}+69\bar{v}^{4}-45\bar{v}^{6}}{\bar{v}}+3\frac{(1-\bar{v}^{2})^{2}(1-2\bar{v}^{2}-15\bar{v}^{4})}{\bar{v}^{2}}\log\left(\frac{1-\bar{v}}{1+\bar{v}}\right).\label{As}
\end{equation}

For the two-bubble contribution, the first term in Eq.~(\ref{eItotHF})
will not vanish (except for $\bar{v}=1$; see below), so we keep only
this term. To lowest order in $\epsilon$, we have 
\begin{align}
\Delta_{HF}^{(d)} & =\frac{\pi\epsilon^{-5}\omega_*^2}{24}\int_{-\infty}^{\infty}d \bar t \int_{-\infty}^{\bar t} dt_{N}\Gamma(t_{N})\int_{-\infty}^{\bar t}dt_{N}'\Gamma(t_{N}')\nonumber \\
 & \times\int_{0}^{\infty}dy\cos y \int_{\bar{v}y}^{\infty}\frac{dx}{x^{4}} e^{-I(\bar t)} \frac{j_{2}(x)}{x^{2}}Q_{+}(s,R,R_{-})Q_{-}(s,R',R_{-}),\label{Deltaenvd-1}
\end{align}
with
\begin{equation}
Q_{+}=-8\epsilon^{3}\bar{R}^{3}\bar{v}y(x^{2}-\bar{v}^{2}y^{2}),\quad Q_{-}=8\epsilon^{3}\bar{R}^{\prime3}\bar{v}y(x^{2}-\bar{v}^{2}y^{2}),
\end{equation}
 which give again an overall factor of $\epsilon=\omega^{-1}$. The
integrals with respect to $t_{N}$ and $t_{N}'$ affect only the factors
$\bar{R}^{3},\bar{R}^{\prime3}$, and give factors $I_{3}(\bar t)$.
The integrations with respect to $x$ and $y$ can be done analytically
again (it is convenient to interchange them), and we obtain
\begin{equation}
\Delta_{HF}^{(d)}=\omega^{-1} \, \frac{\pi \omega_{*}^{2}}{18} \int_{-\infty}^{\infty} d \bar t \, 
e^{-\frac{4\pi}{3}I_{3}(\bar t)} I_{3}(\bar t)^{2}A^{(d)}(\bar{v}),\label{HFAd}
\end{equation}
with 
\begin{equation}
A^{(d)}(\bar{v})=-(1-\bar{v}^{2})\left[2\frac{3+4\bar{v}^{2}-15\bar{v}^{4}}{\bar{v}}-3\frac{(1+\bar{v}^{2}+3\bar{v}^{4}-5\bar{v}^{6})}{\bar{v}^{2}}\log\left(\frac{1+\bar{v}}{1-\bar{v}}\right)\right].\label{Ad}
\end{equation}
Notice that this contribution vanishes for $\bar{v}=1$. Therefore,
in the ultra-relativistic limit, the two-bubble contribution falls
like $\omega^{-2}$, as observed in the computations of Ref.~\cite{jt17}.

This approximation for high frequencies is useful since in this limit
the integrals in (\ref{Deltaenvs})-(\ref{Deltaenvd}) become difficult
to compute numerically due to the highly-oscillatory integrand. We
have found only the leading term, but higher orders can be obtained
in the same way. To calculate the integrals $I_{i}(\bar t)$
and the final integral with respect to $\bar t$ in Eqs.~(\ref{HFAs}) and (\ref{HFAd}),
we need to know the nucleation rate $\Gamma(t)$ as well as the wall
velocity $v(t)$.

\subsection{Constant velocity}

In the case of a constant wall velocity we have $R=v(t-t_{N})$, $R'=v(t'-t_{N}')$,
$R_{-}=vt_{-}$, $R_{+}=v(t_{+}-t_{N}-t_{N}')$, and the expressions
 simplify significantly.
For a nucleation rate of the form (\ref{gammaadim}), it is convenient to use $ \omega_*=\omega_b=t_b^{-1} $ as the reference frequency, and to use the dimensionless variables $\tau=(t-t_{*})/t_{b}$, $\tau'=(t'-t_{*})/t_{b}$. Thus,
we shall make the change of variables 
\begin{equation}
	\tau_{-}=\frac{t_{-}}{t_{b}},\;\tau_{+}=\frac{t_{+}-2t_{*}}{t_{b}},\;\tau_{N}=\frac{t_{N}-t_{*}}{t_{b}},\:\tau_{N}'=\frac{t_{N}'-t_{*}}{t_{b}},\:\tau_{s}=\frac{s}{vt_{b}}\label{cambiovar}
\end{equation}
in the integrals of Eqs.~(\ref{Bgrals})-(\ref{Bgrald}), and
$ \bar \tau =\bar t/t_b$ in Eqs.~(\ref{HFAs}) and (\ref{HFAd}).

\subsubsection{Low frequency}

In the
single-bubble case, the polynomials $P_{i}$ given by Eqs.~(\ref{P0})-(\ref{P2})
are homogeneous functions of degree 8, so we have
\begin{equation}
P_{i}(R_{+},R_{-},s)=(vt_{b})^{8}P_{i}(\tau_{+}-2\tau_{N},\tau_{-},\tau_{s}).\label{Pihomog}
\end{equation}
Changing the order of integration with respect to $\tau_{N}$ and
$\tau_{s}$, we obtain 
\begin{equation}
B^{(s)}=\frac{v^{3}t_{b}^{3}}{48}\int_{-\infty}^{\infty}d\tau_{+}\int_{0}^{\infty}d\tau_{-}\int_{\tau_{-}}^{\infty}\frac{d\tau_{s}}{\tau_{s}^{3}}e^{-I_{\mathrm{tot}}}\int_{-\infty}^{\frac{\tau_{+}-\tau_{s}}{2}}d\tau_{N}\,\tilde{\Gamma}(\tau_{N})\left[P_{0}+\frac{P_{1}}{3}+\frac{P_{2}}{15}\right],\label{Bs}
\end{equation}
with $P_{i}$ evaluated at the dimensionless variables as in the right-hand
side of Eq.~(\ref{Pihomog}). For the two-bubble case, the polynomials
$Q_{\pm}$, Eqs.~(\ref{Qmas})-(\ref{Qmenos}), are homogeneous
functions of degree 6. Changing the order of integration, we obtain 
\begin{equation}
B^{(d)}=\frac{\pi v^{3}t_{b}^{3}}{48}\int_{-\infty}^{\infty}\negmedspace d\tau_{+}\int_{0}^{\infty}\negmedspace d\tau_{-}\int_{\tau_{-}}^{\infty}\frac{d\tau_{s}}{\tau_{s}^{4}}\frac{e^{-I_{\mathrm{tot}}}}{15}\int_{-\infty}^{\frac{\tau_{+}-\tau_{s}}{2}}\negmedspace d\tau_{N}\tilde{\Gamma}(\tau_{N})Q_{+}\int_{-\infty}^{\frac{\tau_{+}-\tau_{s}}{2}}\negmedspace d\tau_{N}'\tilde{\Gamma}(\tau_{N}')Q_{-},\label{Bd}
\end{equation}
where the quantities $Q_{\pm}$ are now given by 
\begin{align}
Q_{+} & (\tau_{s},\tau-\tau_{N},\tau_{-})=(\tau_{s}^{2}-\tau_{-}^{2})\left[(\tau_{+}-2\tau_{N})^{2}-\tau_{s}^{2}\right]\left[\tau_{s}^{2}-\tau_{-}(\tau_{+}-2\tau_{N})\right],\label{Qmasadim}\\
Q_{-} & (\tau_{s},\tau'-\tau_{N}',\tau_{-})=(\tau_{s}^{2}-\tau_{-}^{2})\left[(\tau_{+}-2\tau_{N}')^{2}-\tau_{s}^{2}\right]\left[\tau_{s}^{2}+\tau_{-}(\tau_{+}-2\tau_{N}')\right].\label{Qmenadim}
\end{align}
Finally, for the dimensionless quantities appearing in $I_{\mathrm{tot}}$ we have
\begin{equation}
I=\frac{4\pi}{3}\int_{-\infty}^{\tau}d\tau''\tilde{\Gamma}(\tau'')(\tau-\tau'')^{3}\label{Itilde}
\end{equation}
and 
\begin{equation}
I_{\cap}=\frac{\pi}{12}\int_{-\infty}^{\frac{\tau_{+}-\tau_{s}}{2}}d\tau''\tilde{\Gamma}(\tau'')(\tau_{+}-2\tau''-\tau_{s})^{2}\left[\tau_{s}+2(\tau_{+}-2\tau'')-\frac{3\tau_{-}^{2}}{\tau_{s}}\right].\label{Iinttil}
\end{equation}
Thus,  Eqs.~(\ref{DeltaLFgral})-(\ref{Bgrald}) give 
\begin{equation}
\Delta_{LF}=D\,v^{3}\,\omega^{3}/\omega_{b}^{3},\label{Deltalow}
\end{equation}
with a numerical coefficient $D=(D^{(s)}+\pi D^{(d)})/48$,
where $D^{(s)}$ and $D^{(d)}$ are given by the integrals in Eqs.~(\ref{Bs})
and (\ref{Bd}), respectively. By definition, the dimensionless function
$\tilde{\Gamma}(\tau)$ does not depend on $t_{b}$ or $v$, so the
parametric dependence is $\Delta_{LF}\propto v^{3}t_{b}^{3}\omega^3$.

\subsubsection{High frequency}

For constant velocity, we have $\bar{v}(\bar t)=v$, and the functions
$A^{(s)}(v)$ and $A^{(d)}(v)$
do not depend on the integration variable $\bar t$  in Eqs.~(\ref{HFAs}) and (\ref{HFAd}). Using again the
dimensionless form of the nucleation rate, Eq.~(\ref{In}) becomes
\begin{equation}
I_{n}(\bar t) = (vt_{b})^{n-3}\int_{-\infty}^{\bar \tau} d\tau''\,\tilde{\Gamma}(\tau'') \left(\bar \tau-\tau''\right)^{n}
\equiv(vt_{b})^{n-3}\tilde{I}_{n}(\bar \tau)\label{Itilden}
\end{equation}
Thus, we have 
\begin{equation}
\Delta_{HF}=A(v)\,\omega_{b}/\omega,\label{Deltahigh}
\end{equation}
with 
\begin{equation}
A(v)=C^{(s)}A^{(s)}(v)+C^{(d)}A^{(d)}(v),
\label{Av}
\end{equation}
where the two numerical coefficients $C^{(s)}$ and $C^{(d)}$ are
given by
\begin{equation}
C^{(s)}=\frac{\pi}{72}\int_{-\infty}^{\infty}\negmedspace d \bar \tau e^{-\frac{4\pi}{3}\tilde{I}_{3}(\bar\tau)} \tilde{I}_{4}(\bar\tau) \tilde{I}_{2}(\bar\tau) , \; C^{(d)}=\frac{\pi}{18}\int_{-\infty}^{\infty}\negmedspace d \bar\tau e^{-\frac{4\pi}{3}\tilde{I}_{3}(\bar\tau)} [\tilde{I}_{3}(\bar\tau)]^{2},
\label{Cs}
\end{equation}
and we remark that the functions $A^{(s)}(v)$ and $ A^{(d)}(v) $ are given analytically by Eqs.~(\ref{As}) and (\ref{Ad}), respectively.
For most of the nucleation rates considered below, the coefficients $ C^{(s)} $ and $ C^{(d)} $ can also be calculated analytically.

\subsubsection{Interpolation}

Although we cannot give an analytic fit for the spectrum for an arbitrary
nucleation rate, we note that, from the two asymptotes (\ref{Deltalow})
and (\ref{Deltahigh}), we may obtain a rough approximation for the
whole spectrum. The intersection of the curves of $\Delta_{LF}$ and
$\Delta_{HF}$ occurs at $\omega=\omega_{\times}$, $\Delta=\Delta_{\times}$,
with 
\begin{equation}
\omega_{\times}/\omega_{b}=\left[A(v)/Dv^{3}\right]^{1/4},\quad\Delta_{\times}=\left[Dv^{3}A(v)^{3}\right]^{1/4}.\label{wcruce}
\end{equation}
We shall check with specific examples below that these values give
the approximate position of the peak frequency $\omega_{p}$ as well
as an order-of-magnitude estimate of the amplitude $\Delta(\omega_{p})$.
The actual value of the latter is below the intersection point, and
a better approximation to $\Delta(\omega)$ is given by the simple
interpolation 
\begin{equation}
\Delta_{\mathrm{int}}(\omega)=\left(\Delta_{LF}^{-1/2}+\Delta_{HF}^{-1/2}\right)^{-2}.\label{interp}
\end{equation}
The maximum of Eq.~(\ref{interp}) is at $\sqrt{3}\omega_{\times}$,
and the value of $\Delta_{\mathrm{int}}$ at this frequency is given
by $3^{3/2}\Delta_{\times}/16$.
For small velocity we have 
\begin{equation}
\frac{\omega_{\times}}{\omega_b} = 4\left(\frac{C^{(d)}+4C^{(s)}}{5D}\right)^{\frac14}+\mathcal{O}\left(v^{2}\right),
\quad \Delta_{\times}=64D^{\frac14}\left(\frac{C^{(d)}+4C^{(s)}}{5}\right)^{\frac34}v^{3}+\mathcal{O}\left(v^{5}\right). 	
\label{wxvchica}
\end{equation}
Therefore, the peak frequency is approximately fixed for most of the velocity range, 
while the amplitude is roughly proportional to $v^3$.
Near $v=1$ the intersection point departs from Eq.~(\ref{wxvchica}). We have 
\begin{equation}
\frac{\omega_{\times}}{\omega_b}=2\left(\frac{2C^{(s)}}{D}\right)^{1/4}+\mathcal{O}\left(1-v\right), \quad
\Delta_{\times}=8D^{1/4}\left(2C^{(s)}\right)^{3/4}+\mathcal{O}\left(1-v\right).
\label{wxv1}
\end{equation}

\section{Specific examples}

\label{specific}

We shall now calculate the GW spectrum for a few specific cases. 
We begin by writing down the expressions for the case of a constant wall velocity.
Notice that the expressions for the complete
spectrum $\Delta$, Eqs.~(\ref{Deltaenvs})-(\ref{Deltaenvd}), are very similar to those for the low-frequency
limit, Eqs.~(\ref{Bgrals})-(\ref{Bgrald}). 
For a constant wall velocity it will be useful to do the same 
change of variables we used in the previous section, Eq.~(\ref{cambiovar}), 
and we obtain expressions 
for $\Delta^{(s)}$ and $\Delta^{(d)}$ which are similar to those for $ B^{(s)} $ and $ B^{(d)} $,
Eqs.~(\ref{Bs}) and (\ref{Bd}), but including the factor $\omega^{3}$
and the oscillating functions shown in Eq.~(\ref{depwLF}). Defining
$\tilde{\omega}\equiv\omega/\omega_{b}$, we have 
\begin{equation}
\Delta^{(s)}=\frac{v^{3}\tilde{\omega}^{3}}{48}\int_{-\infty}^{\infty}d\tau_{+}\int_{0}^{\infty}d\tau_{-}\cos(\tilde{\omega}\tau_{-})\int_{\tau_{-}}^{\infty}\frac{d\tau_{s}}{\tau_{s}^{3}}\sum_{i=0}^{2}\frac{j_{i}(v\tilde{\omega}\tau_{s})}{(v\tilde{\omega}\tau_{s})^{i}}F_{i}(\tau_{+},\tau_{-},\tau_{s})e^{-I_{\mathrm{tot}}},
\label{Dsvconst}
\end{equation}
 where
\begin{equation}
F_{i}=\int_{-\infty}^{\frac{\tau_{+}-\tau_{s}}{2}}d\tau_{N}\,\tilde{\Gamma}(\tau_{N})P_{i}(\tau_{+}-2\tau_{N},\tau_{-},\tau_{s}),\label{F}
\end{equation}
with the polynomials $P_{i}$ defined in Eqs.~(\ref{P0})-(\ref{P2}),
and
\begin{equation}
\Delta^{(d)}=\frac{\pi v^{3}\tilde{\omega}^{3}}{48}\int_{-\infty}^{\infty}d\tau_{+}\int_{0}^{\infty}d\tau_{-}\cos(\tilde{\omega}\tau_{-})\int_{\tau_{-}}^{\infty}\frac{d\tau_{s}}{\tau_{s}^{4}}\frac{j_{2}(v\tilde{\omega}\tau_{s})}{(v\tilde{\omega}\tau_{s})^{2}}G_{+}G_{-}e^{-I_{\mathrm{tot}}},\label{Ddvconst}
\end{equation}
where 
\begin{equation}
G_{\pm}(\tau_{+},\tau_{-},\tau_{s})=
\int_{-\infty}^{\frac{\tau_{+}-\tau_{s}}{2}}d\tau_{N}\,\tilde{\Gamma}(\tau_{N})
Q_{\pm}(\tau_{s},{\textstyle \frac{\tau_{+}\mp\tau_{-}}{2}}-\tau_{N},\tau_{-}).\label{G}
\end{equation}
The expressions for the quantities $Q_{\pm}$ in terms of these variables
are given in Eqs.~(\ref{Qmasadim})-(\ref{Qmenadim}). The quantity
$I_{\mathrm{tot}}$ is given by Eqs.~(\ref{Itilde})-(\ref{Iinttil})
as a function of $\tau_{+},$$\tau_{-}$ and $\tau_{s}$. 

\subsection{Exponential nucleation rate}

The case of an exponential nucleation rate (and a constant wall velocity)
was studied numerically in Refs.~\cite{kt93,hk08,w16} and analytically
in Ref.~\cite{jt17}. We shall now see that Eqs.~(\ref{Dsvconst})-(\ref{G})
give for this case the analytic result of Ref.~\cite{jt17}. We use
the parametrization (\ref{nuclexpadim}) for the nucleation rate,
which is of the form (\ref{gammaadim}) with $\tilde{\Gamma}(\tau)=\frac{1}{8\pi}e^{\tau}$,
$t_{b}=\beta^{-1}$, and $v_{b}=v$. In this case, we have $\omega_{b}=\beta$,
and the dimensionless spectrum (\ref{Deltadef}) (with $ \omega_* =\omega_b $) coincides with the
expression (\ref{Deltadef-exp}). Since $\tilde{\Gamma}$ is an exponential
and $P_{i},Q_{\pm}$ are polynomials, the integrals (\ref{F}) and
(\ref{G}) are straightforward. We obtain 
\begin{equation}
F_{i}=\frac{2}{\pi}e^{\tau_{+}/2}e^{-\tau_{s}/2}\tilde{F}_{i}(\tau_{-},\tau_{s}),\quad G_{+}G_{-}=\frac{1}{4\pi^{2}}e^{\tau_{+}}e^{-\tau_{s}}\tilde{G}(\tau_{-},\tau_{s})\tilde{G}(-\tau_{-},\tau_{s}),
\label{FGfinal}
\end{equation}
with
\begin{align}
\tilde{F}_{0}= & \,2\left(\tau_{s}^{2}+6\tau_{s}+12\right)\left(\tau_{s}^{2}-\tau_{-}^{2}\right)^{2}, \label{Ftil0} \\
\tilde{F}_{1}= & \,2\left(\tau_{s}^{2}-\tau_{-}^{2}\right)\left[(\tau_{s}^{3}+4\tau_{s}^{2}+12\tau_{s}+24)\tau_{s}^{2}-\left(\tau_{s}^{3}+12\tau_{s}^{2}+60\tau_{s}+120\right)\tau_{-}^{2}\right],\\
\tilde{F}_{2}= & \,\frac{1}{2}\left[(\tau_{s}^{4}+4\tau_{s}^{3}+20\tau_{s}^{2}+72\tau_{s}+144)\tau_{s}^{4}+\left(\tau_{s}^{4}+20\tau_{s}^{3}+180\tau_{s}^{2}+840\tau_{s}+1680\right)\tau_{-}^{4}\right.\nonumber \\
 & \left.-\left(2\tau_{s}^{4}+24\tau_{s}^{3}+168\tau_{s}^{2}+720\tau_{s}+1440\right)\tau_{s}^{2}\tau_{-}^{2}\right],\\
\tilde{G}= & \,(\tau_{s}^{2}-\tau_{-}^{2})\left[\tau_{s}^{3}+2\tau_{s}^{2}-\tau_{-}(\tau_{s}^{2}+6\tau_{s}+12)\right]. \label{Gtil}
\end{align}
On the other hand, we have $I(t)=e^{\tau}=e^{(\tau_{+}-\tau_{-})/2}$,
$I(t')=e^{\tau'}=e^{(\tau_{+}+\tau_{-})/2}$, and $I_{\cap}=\frac{1}{4}e^{\tau_{+}/2}e^{-\tau_{s}/2}\left(\tau_{s}+4-\tau_{-}^{2}/\tau_{s}\right)$.
Summing these contributions, we obtain
\begin{equation}
I_{\mathrm{tot}}=e^{\tau_{+}/2}\left[2\cosh\left(\tau_{-}/2\right)-e^{-\tau_{s}/2}\left(\tau_{s}/4+1-\tau_{-}^{2}/4\tau_{s}\right)\right].\label{Itot}
\end{equation}
Notice that, in all these expressions, the variable $\tau_{+}$ appears
only in exponentials $e^{\tau_{+}/2},e^{\tau_{+}}$, and the integration
with respect to this variable can be readily done using the substitution
$u=e^{\tau_{+}/2}$. We obtain 
\begin{equation}
\Delta^{(s)}=\frac{(v\tilde{\omega})^{3}}{12\pi}\int_{0}^{\infty}d\tau_{-}\cos(\tilde{\omega}\tau_{-})\int_{\tau_{-}}^{\infty}\frac{d\tau_{s}}{\tau_{s}^{3}}\frac{\tilde{F}_{0}j_{0}(v\tilde{\omega}\tau_{s})+\tilde{F}_{1}\frac{j_{1}(v\tilde{\omega}\tau_{s})}{v\tilde{\omega}\tau_{s}}+\tilde{F}_{2}\frac{j_{2}(v\tilde{\omega}\tau_{s})}{(v\tilde{\omega}\tau_{s})^{2}}}{2\cosh\left(\frac{\tau_{-}}{2}\right)e^{\frac{\tau_{s}}{2}}-1-(\tau_{s}^{2}-\tau_{-}^{2})/4\tau_{s}},\label{deltascolexpfinal}
\end{equation}
\begin{equation}
\Delta^{(d)}=\frac{(v\tilde{\omega})^{3}}{96\pi}\int_{0}^{\infty}d\tau_{-}\cos(\tilde{\omega}\tau_{-})\int_{\tau_{-}}^{\infty}\frac{d\tau_{s}}{\tau_{s}^{4}}\frac{\tilde{G}(\tau_{-},\tau_{s})\tilde{G}(-\tau_{-},\tau_{s})\frac{j_{2}(v\tilde{\omega}\tau_{s})}{(v\tilde{\omega}\tau_{s})^{2}}}{\left[2\cosh\left(\frac{\tau_{-}}{2}\right)e^{\frac{\tau_{s}}{2}}-1-(\tau_{s}^{2}-\tau_{-}^{2})/4\tau_{s}\right]^{2}},\label{deltadcolexpfinal}
\end{equation}
in agreement with Ref.~\cite{jt17}.

\begin{figure}[tb]
\centering
\includegraphics[width=0.7\textwidth]{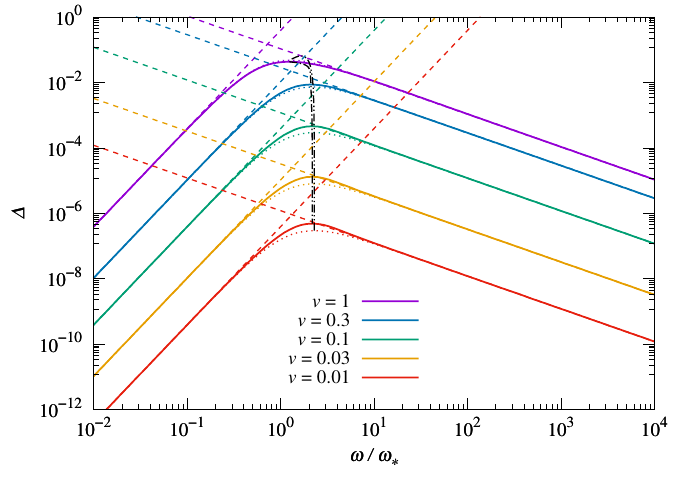}
\caption{The GW spectrum for the exponential nucleation rate, with $\omega_{*}=\beta$
(solid lines). The dashed lines indicate the asymptotes, and the dotted
lines correspond to the interpolation (\ref{interp}). The dash-dot
lines indicate the peak values $(\omega_{p},\Delta_{p})$ for the
spectrum, and dash-dot-dot lines those for the interpolation.
\label{figdeltaexp}}
\end{figure}

In Fig.~\ref{figdeltaexp} we plot the spectrum for several wall
velocities, as well as the low-frequency and high-frequency approximations
$\Delta_{LF},\Delta_{HF}$ given by Eqs.~(\ref{Deltalow}) and (\ref{Deltahigh}),
respectively.
The coefficients for these approximations 
are\footnote{The functions $ \tilde I_n(\bar\tau) $ defined in Eq.~(\ref{Itilden}) are given by 
$ \tilde I_2= \frac{1}{4\pi}e^{\bar \tau}$, $ \tilde I_3= \frac{3}{4\pi}e^{\bar\tau}$, 
and $ \tilde I_4= \frac{3}{\pi}e^{\bar\tau}$.} 
$D\simeq0.3820$, $C^{(s)}=(96\pi)^{-1}$,
and $C^{(d)}=(32\pi)^{-1}$. We see that the asymptotic curves give
an order-of-magnitude approximation in the whole range. The figure
also shows the simple interpolation (\ref{interp}). Its maximum gives
a good approximation for the peak frequency, $\omega_{p}\simeq\sqrt{3}{\omega}_{\times}$.
The error is less than a 10\% for all the curves. On the other hand,
the maximum value of $\Delta$ for the interpolation gives a rough
approximation for the peak amplitude, $\Delta_{p}\simeq3^{3/2}\Delta_{\times}/16$,
although for some of the curves this value departs more than a 50\%
from the actual value.

\subsection{Simultaneous nucleation}

We now calculate the GW spectrum for a delta-function nucleation rate.
This case was considered in Ref.~\cite{jlst17} as a limit of a Gaussian
nucleation rate. In that work, the general expressions (B.22) and
(B.28) have the same form of our Eqs.~(\ref{Dsvconst})-(\ref{Ddvconst}).
However, their specific expressions for the integrands, Eqs.~(B.30)-(B.36),
are somewhat cumbersome for a direct comparison. On the other hand,
we shall perform the integral with respect to $\tau_{-}$ analytically,
which greatly simplifies the remaining numerical integrations and
will allow us to consider a much wider frequency range. The particular case $v=1$
was considered also in Ref.~\cite{w16}. In appendix \ref{shapes}
we compare the different numerical results.

We use the parametrization (\ref{nuclsimultadim}) of the nucleation
rate, for which the dimensionless rate is $\tilde{\Gamma}(\tau)=\delta(\tau)$
and the time scale is defined as $t_{b}=d_{b}/v$. The associated frequency
is $\omega_{b}=t_{b}^{-1}=v/d_{b}$, and we shall use this as the reference frequency $ \omega_* $ for the dimensionless spectrum $ \Delta $. 
Due to the delta function,
the integrals (\ref{F}) and (\ref{G}) are trivial, and we obtain
\begin{equation}
F_{i}=P_{i}(\tau_{+},\tau_{-},\tau_{s})\Theta(\tau_{+}-\tau_{s}),\quad G_{-}G_{+}=Q(\tau_{+},\tau_{-},\tau_{s})\Theta(\tau_{+}-\tau_{s}),
\end{equation}
where $P_{i}$ are the polynomials defined in Eqs.~(\ref{P0})-(\ref{P2}),
and 
\begin{equation}
Q=(\tau_{+}^{2}-\tau_{s}^{2})^{2}(\tau_{s}^{2}-\tau_{-}^{2})^{2}(\tau_{s}^{4}-\tau_{-}^{2}\tau_{+}^{2}).\label{Q}
\end{equation}
Hence, Eqs.~(\ref{Dsvconst}) and (\ref{Ddvconst}) become 
\begin{equation}
\Delta^{(s)}=\frac{v^{3}\tilde{\omega}^{3}}{48}\int_{0}^{\infty}d\tau_{+}\int_{0}^{\tau_{+}}d\tau_{-}\cos(\tilde{\omega}\tau_{-})\int_{\tau_{-}}^{\tau_{+}}\frac{d\tau_{s}}{\tau_{s}^{3}}\sum_{i=0}^{2}\frac{j_{i}(v\tilde{\omega}\tau_{s})}{(v\tilde{\omega}\tau_{s})^{i}}P_{i}(\tau_{+},\tau_{-},\tau_{s})e^{-I_{\mathrm{tot}}},\label{Deltassimult}
\end{equation}
\begin{equation}
\Delta^{(d)}=\frac{\pi v^{3}\tilde{\omega}^{3}}{48}\int_{0}^{\infty}d\tau_{+}\int_{0}^{\tau_{+}}d\tau_{-}\cos(\tilde{\omega}\tau_{-})\int_{\tau_{-}}^{\tau_{+}}\frac{d\tau_{s}}{\tau_{s}^{4}}\frac{j_{2}(v\tilde{\omega}\tau_{s})}{(v\tilde{\omega}\tau_{s})^{2}}Q(\tau_{+},\tau_{-},\tau_{s})e^{-I_{\mathrm{tot}}},\label{Deltadsimult}
\end{equation}
The integrals for $I_{\mathrm{tot}}$ are also trivial, and we obtain
\begin{equation}
I_{\mathrm{tot}}=\frac{\pi}{12}\left(2\tau_{+}^{3}-\tau_{s}^{3}+3\tau_{+}^{2}\tau_{s}\right)+\frac{\pi}{4}\frac{(\tau_{+}+\tau_{s})^{2}}{\tau_{s}}\tau_{-}^{2}\label{Itotsimult}
\end{equation}
Since the exponent (\ref{Itotsimult}) is quadratic in $\tau_{-}$,
the integrals with respect to this variable can be calculated analytically.
We must first interchange the integrals with respect to $\tau_{-}$
and $\tau_{s}$ using $\int_{0}^{\tau_{+}}d\tau_{-}\int_{\tau_{-}}^{\tau_{+}}d\tau_{s}=\int_{0}^{\tau_{+}}d\tau_{s}\int_{0}^{\tau_{s}}d\tau_{-}$.
We obtain
\begin{equation}
\Delta^{(s)}=\frac{(v\tilde{\omega})^{3}}{48}\int_{0}^{\infty}d\tau_{+}\int_{0}^{\tau_{+}}\frac{d\tau_{s}}{\tau_{s}^{3}}e^{-\frac{\pi}{12}\left(2\tau_{+}^{3}-\tau_{s}^{3}+3\tau_{+}^{2}\tau_{s}\right)}\sum_{i=0}^{2}\frac{j_{i}(v\tilde{\omega}\tau_{s})}{(v\tilde{\omega}\tau_{s})^{i}}\tilde{P}_{i}(\tau_{+},\tau_{s},\tilde{\omega}),
\label{Deltassimultfinal}
\end{equation}
with
\begin{equation}
\tilde{P}_{i}(\tau_{+},\tau_{s},\tilde{\omega})=\int_{0}^{\tau_{s}}d\tau_{-}\cos(\tilde{\omega}\tau_{-})P_{i}(\tau_{+},\tau_{-},\tau_{s})e^{-\frac{\pi}{4}\frac{(\tau_{+}+\tau_{s})^{2}}{\tau_{s}}\tau_{-}^{2}},
\end{equation}
and 
\begin{equation}
\Delta^{(d)}=\frac{\pi(v\tilde{\omega})^{3}}{48}\int_{0}^{\infty}d\tau_{+}\int_{0}^{\tau_{+}}\frac{d\tau_{s}}{\tau_{s}^{4}}e^{-\frac{\pi}{12}\left(2\tau_{+}^{3}-\tau_{s}^{3}+3\tau_{+}^{2}\tau_{s}\right)}\frac{j_{2}(v\tilde{\omega}\tau_{s})}{(v\tilde{\omega}\tau_{s})^{2}}\tilde{Q}(\tau_{+},\tau_{s},\tilde{\omega}),
\label{Deltadsimultfinal}
\end{equation}
with 
\begin{equation}
\tilde{Q}(\tau_{+},\tau_{s},\tilde{\omega})=\int_{0}^{\tau_{s}}d\tau_{-}\cos(\tilde{\omega}\tau_{-})Q(\tau_{+},\tau_{-},\tau_{s})e^{-\frac{\pi}{4}\frac{(\tau_{+}+\tau_{s})^{2}}{\tau_{s}}\tau_{-}^{2}}.
\end{equation}
The analytic expressions for the functions $\tilde{P}_{i}$ and $\tilde{Q}$
are given in appendix \ref{apsimult}.

\begin{figure}[tb]
\centering
\includegraphics[width=0.7\textwidth]{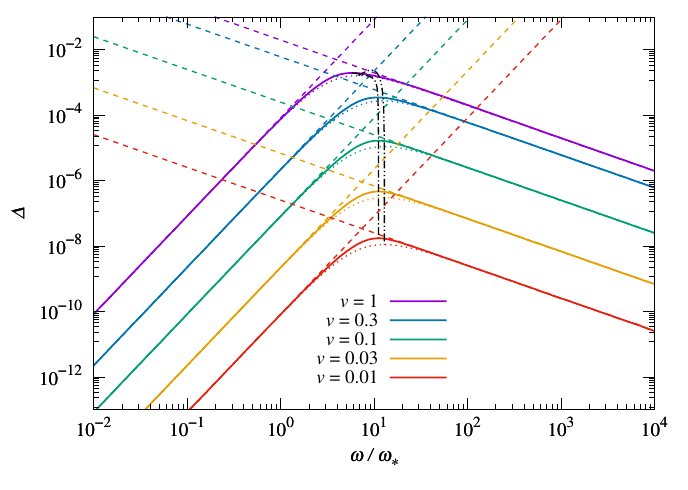}
\caption{Like Fig.~\ref{figdeltaexp} but for the delta-function rate, with
$\omega_{*}=v/d_{b}$. 
\label{figdeltasimult}}
\end{figure}

In appendix \ref{apsimult} we plot separately the single-bubble and
the two-bubble contributions to the GW spectrum. In Fig.~\ref{figdeltasimult}
we plot the complete spectrum together with the asympotic curves and
the interpolation. 
The coefficient of the low frequency approximation is $D\simeq8.066\times10^{-5}$,
while those of the high-frequency
approximation are given by\footnote{The functions 
	$ \tilde I_n(\bar \tau) $ which appear in the expressions for $ C^{(s)} $  and $ C^{(d)} $ are given by 
	$ \tilde I_n= \bar \tau ^n$.} 
\begin{equation}
C^{(s)}=\frac{\Gamma\left(7/3\right)}{128\pi^{4/3}6^{2/3}}\simeq6.123\times10^{-4},\quad C^{(d)}=\frac{\Gamma\left(7/3\right)}{32\pi^{4/3}6^{2/3}}\simeq2.449\times10^{-3}.
\end{equation}
We see that, at the maximum, the interpolations are not as good approximations
as in the previous case. The peak frequency departs more than a 50\%
and the amplitude departs by a factor of 2 in some cases.

\subsection{Gaussian nucleation rate}
\label{GWgau}

We shall now consider a nucleation rate of the form $ e^{-\gamma^{2}(t-t_{m})^{2}} $.
A Gaussian nucleation rate was considered in Ref.~\cite{jlst17} with a different parametrization, 
namely, $e^{\beta (t-t_*)-\gamma^{2}(t-t_*)^{2}} $. 
This parametrization 
is useful when the phase transition occurs away from the maximum of the Gaussian
(in particular, it allows to consider the case $ \gamma=0 $), while 
we are more interested in the case in which the phase transition occurs around this maximum.
For a given physical model, the two exponents correspond to the expansion of $ S(t) $ around two different times
$ t_m , t_* $.
Therefore, the value of $ \gamma $ is different in each case.
We compare the two approaches in more detail in App.~\ref{apgauss}. 

We shall use the parametrization (\ref{gammagauss}), which
is of the form (\ref{gammaadim}), with $ t_b $ replaced by the parameter $ t_{\min} $ 
for simplicity of the expressions.
The dimensionless rate is
$ \tilde \Gamma(\tau)=({\alpha}/{\sqrt{\pi}}) e^{-(\alpha\tau)^2} $,
where $\alpha= \gamma t_{\min}  $. 
The time scale of the nucleation rate is $ t_\Gamma \sim \gamma^{-1}$, but
the duration of the phase transition, $ t_b $, depends mainly on the value of $ t_{\min} $.
We shall use the reference frequency $ \omega_*=t_{\min}^{-1} $, so the dimensionless spectrum is given by
Eqs.~(\ref{Dsvconst})-(\ref{G}), with $ \tilde \omega =\omega/t_{\min}^{-1}$.

\begin{figure}[tb]
	\centering
	\includegraphics[width=0.7\textwidth]{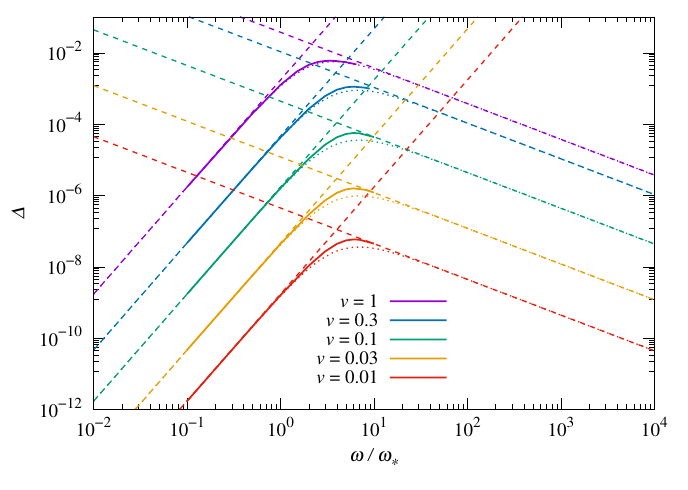}
	\caption{Like Fig.~\ref{figdeltaexp} but for a Gaussian nucleation rate with $ \alpha=1 $. Here, 
		$\omega_{*}=t_{\min}^{-1}$. 
		\label{figdeltagauss}}
\end{figure}

The integrals (\ref{F}) and (\ref{G}) can be done analytically. 
The expressions for $ F_i $ and $ G_\pm $ contain polynomials, the Gaussian function, 
and the error function. These expressions are rather cumbersome, and we write them down in App.~\ref{apgauss}. 
The function $ I_\mathrm{tot} $ also contains error functions which depend on the variables $ t_-,t_+,t_s $,  
so the remaining integrals in Eqs.~(\ref{Dsvconst}) and (\ref{Ddvconst}) cannot be done analytically.
The multiple integration is difficult to do numerically, 
and in Ref.~\cite{jlst17} a limited frequency range around the peak of the spectrum was considered.
In particular, the high frequency behavior cannot be seen in those results.
Therefore, this case provides an example of the usefulness of our asymptotic approximations.
In Fig.~\ref{figdeltagauss} we show the spectrum, the asymptotes, and the interpolation for the case $ \alpha=1 $
and for several values of the wall velocity.
We give the details of the calculation in App.~\ref{apgauss}.
We have computed the exact GW spectrum only in the range $ 0.1\leq\omega/\omega_*\lesssim 10 $. 
For higher frequencies, the multiple integration becomes very difficult due to the highly oscillating integrand.
Nevertheless, we see that in this case the high-frequency asymptote or the interpolation become good approximations.
For lower frequencies, the numerical integration does not present critical difficulties.
In any case, we see that for $ \omega/\omega_* \sim 10^{-1}$ the low-frequency asymptote is already a very good approximation.

\begin{figure}[tb]
	\centering
	\includegraphics[width=0.7\textwidth]{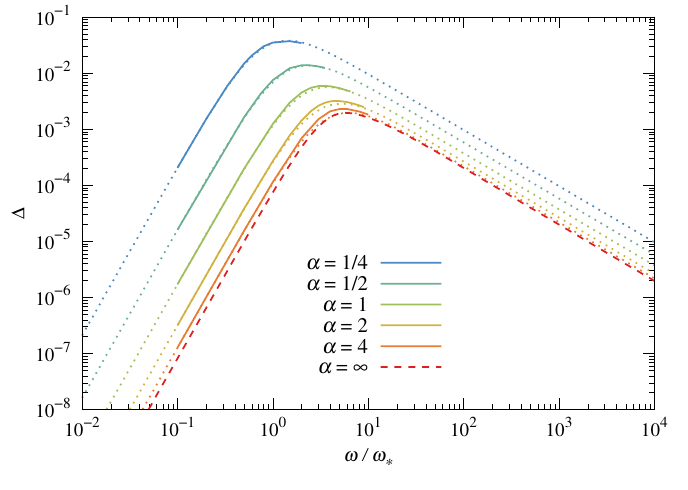}
	\caption{The spectrum (solid lines) and the interpolation approximation (dotted lines) for the Gaussian nucleation rate for  $ v=1 $, with
		$\omega_{*}=t_{\min}^{-1}$. 
		The dashed line corresponds to the delta-function rate. 
		\label{figdeltagaussalfa}}
\end{figure}

In Fig.~\ref{figdeltagaussalfa} we show the GW spectrum  for 
a few values of the parameter $ \alpha $.
The dashed red curve actually corresponds to the simultaneous nucleation considered in the previous subsection.
Indeed, in the limit of large $ \alpha $ we have $ t_\Gamma\ll t_{\min} $, and the Gaussian becomes a delta function.
The opposite limit, $ \alpha\to 0 $, corresponds to
$ t_{\min}\ll t_\Gamma $, but in this case $ t_{\min} $ does not represent the duration of the phase transition, i.e., we have  $ t_b>t_{\min} $. We discuss the dependence with the time scales in the next section.
Notice also that the limit $ \alpha\to 0 $ can be interpreted as $ \gamma\to 0 $.
However, this limit does not coincide with what is usually called a constant nucleation rate. The latter is actually a Heaviside function
since it turns on at a given time $ t_0 $. We consider this case next.

\subsection{Constant nucleation rate}

Although a constant nucleation rate $ \Gamma_0 \Theta(t-t_0) $ is not well motivated physically, we shall discuss it here since it is often used as an approximation
(for its application to the computation of GWs, see \cite{chw18}).
We use the parametrization (\ref{Gammaconst}) for the nucleation rate, so we have $\tilde{\Gamma}(\tau)=\frac{3}{\pi}\Theta(\tau)$,
and 
we compute the dimensionless spectrum (\ref{Deltadef}) with $ \omega_*=\omega_b=t_b^{-1} $, which is given by Eqs.~(\ref{Dsvconst})-(\ref{G}).
In this case, the integrands in Eqs.~(\ref{F}) and (\ref{G}) are polynomials, and we obtain, omitting a Heaviside $ \Theta(\tau_+-\tau_s) $ in the expressions,
\begin{align}
F_{0}= & \, \frac{1}{10\pi}(\tau_{+}-\tau_{s})^{3}(\tau_{-}^{2}-\tau_{s}^{2})^{2}(3\tau_{+}^{2}+9\tau_{+}\tau_{s}+8\tau_{s}^{2}),	
\\
F_{1}= & \, \frac{1}{5\pi}(\tau_{s}^{2}-\tau_{-}^{2})
\left[15\tau_{+}\tau_{-}^{2}(\tau_{+}^{2}-\tau_{s}^{2})^{2}+\tau_{s}^{2}\tau_{+}(45\tau_{s}^{4}-10\tau_{+}^{2}\tau_{s}^{2}-3\tau_{+}^{4})-32\tau_{s}^{7}\right],
\\
F_{2} = & \, \frac{1}{10\pi}
\left[15\tau_{-}^{4}\tau_{+}\left(7\tau_{+}^{4}-10\tau_{+}^{2}\tau_{s}^{2}+3\tau_{s}^{4}\right)
	+30\tau_{-}^{2}\tau_{s}^{2}\tau_{+}\left(\tau_{s}^{4}+2\tau_{+}^{2}\tau_{s}^{2}-3\tau_{+}^{4}\right)\right.
\\
	&	
	\left. +\tau_{s}^{4}\tau_{+}\left(9\tau_{+}^{4}+10\tau_{+}^{2}\tau_{s}^{2}+45\tau_{s}^{4}\right)-64\tau_{s}^{9}\right],
\\
G_{\pm}= & \, \mp\frac{1}{8\pi}(\tau_{+}-\tau_{s})^{2}(\tau_{s}^{2}-\tau_{-}^{2})\left[3\tau_{-}(\tau_{+}+\tau_{s})^{2}\mp 4\tau_{s}^{2}(\tau_{+}+2\tau_{s})\right].
\end{align}
On the other hand, we have $I=\tau^4$, $ I_\cap=\frac{1}{16}(\tau_{+}-\tau_{s})^{3}(\tau_{+}+\tau_{s}-2\tau_{-}^{2}/\tau_{s}) $, and
\begin{equation}
	I_\mathrm{tot}=\frac{1}{16}\left[(\tau_{+}-\tau_{-})^{4}+(\tau_{+}+\tau_{-})^{4}-(\tau_{+}-\tau_{s})^{3}\left(\tau_{+}+\tau_{s}-2\tau_{-}^{2}/\tau_{s}\right)\right]
\end{equation}
The remaining integrals with respect to $ \tau_+ $, $ \tau_- $, and $ \tau_s $ in Eqs.~(\ref{Dsvconst}) and (\ref{Ddvconst}) cannot be done analytically.

\begin{figure}[tb]
	\centering
	\includegraphics[width=0.7\textwidth]{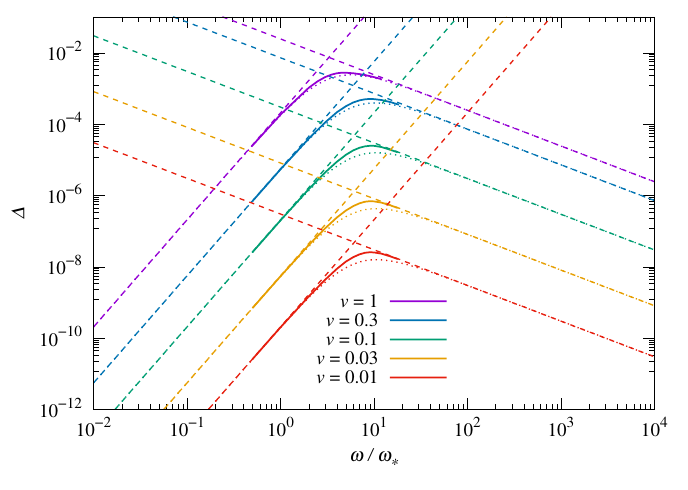}
	\caption{Like Fig.~\ref{figdeltaexp} but for a constant nucleation rate, with $\omega_{*}=t_{b}^{-1}$. 
		\label{figdeltaconst}}
\end{figure}

In Fig.~\ref{figdeltaconst} we show the spectrum, the asymptotes, and the interpolation, for several values of the wall velocity. 	
The coefficient of the low-frequency approximation (\ref{Deltalow}) is given by 
\begin{equation}
	D=\frac{1}{48}\int_{-\infty}^{\infty}d\tau_{+}\int_{0}^{\infty}d\tau_{-}
	\int_{\tau_{-}}^{\tau_+} \frac{d\tau_{s}}{\tau_{s}^{3}}
	e^{-I_{\mathrm{tot}}}	
	\left[F_{0}+\frac{F_{1}}{3}+\frac{F_{2}}{15}+\frac{\pi G_+G_-}{15\tau_s} \right] 
	\simeq 2.046\times 10^{-4}. 
\end{equation}
For the high-frequency approximation (\ref{Deltahigh})-(\ref{Av}), the coefficients are given 
by\footnote{In this case we have
	$ \tilde I_n(\bar\tau) = \frac{3}{(n+1)\pi}\bar\tau^{n+1} $.}
\begin{equation}
	C^{(s)}=\frac{\Gamma \left({9}/{4}\right)}{480 \pi } \simeq 7.513\times10^{-4},
	\quad 
	C^{(d)}=\frac{\Gamma \left({9}/{4}\right)}{128 \pi }  \simeq 2.818\times10^{-3}.
\end{equation}
The case $ v=1 $ can be compared with the lattice simulations of Ref.~\cite{chw18}. 
We find that the peak for envelope approximation is slightly to the left with respect to that computation.
We discuss the differences between these approaches in Sec.~\ref{conclu}.

\section{Time and size scales}
\label{scales}

It has been discussed in the literature whether the GWs should inherit
the characteristic frequency or the characteristic length of the source
(see, e.g., \cite{kkt94,kmk02,dgn02,cd06}). This issue was specifically
analyzed in Ref.~\cite{cds06}. For a spatially homogeneous and short
lived source, the GWs are expected to inherit the characteristic length.
However, for the bubble collision mechanism, it turns out that the characteristic frequency 
is of the order of the time scale $t_b$ rather than the length scale $d_b=vt_b$. 
This was observed numerically for the exponential rate in Refs.~\cite{hk08,jt17}, 
and can be seen in all the plots of the previous section,	
where the peak frequency $ \omega_p $ is around the value $ \omega_* \sim t_b^{-1}$.
Below we discuss this issue in more detail.

\subsection{The bubble size distribution}

Some of the cases considered above correspond to very different
bubble size distributions. For instance, for an exponential nucleation, 
smaller bubbles, which
nucleate later, have exponentially higher number densities than larger bubbles, which nucleate earlier. 
Besides, newer bubbles only nucleate in the increasingly
smaller regions remaining in the false vacuum, so the space distribution
also depends on the bubble size. 
In contrast, for a simultaneous nucleation, all the
bubbles have the same size at any time during the phase transition.
However, 
the shape of the spectrum is very similar for the two nucleation rates. 
This can be seen more clearly in Fig.~\ref{figjapos} in appendix \ref{shapes}. 
Therefore, the presence of different size scales does not seem to be relevant for GW production.

As argued in Ref.~\cite{hk08}, 
the result $ \omega_p\sim t_b^{-1} $ may be explained by the fact that, for $v\ll1$, the duration
of the phase transition is not actually short in comparison to the
scale $ d_b=vt_b $. Hence, the GWs do not inherit the distance scale.
This explains also why the size distribution is not a decisive factor. 

Notice, indeed, that the relevant bubble radius
is \emph{at most} of order $ vt_b $. 
This is quite clear for a simultaneous nucleation, since the bubble radius is limited by the bubble separation $ d_b $
and the time of bubble expansion is $ \sim d_b/v $.
In the exponential case, the average radius at any time is approximately
given by $v\beta^{-1}$, and the width of the radius distribution
is of the same order. 
Since the released energy is proportional to the bubble volume, it is 
sometimes assumed that the volume distribution of bubbles
is the relevant quantity \cite{kt93}. This quantity also has its peak at
a radius of order $v\beta^{-1}$. 

One could argue that, since the walls of different bubbles join to form larger domains,
in the evolution of this system of walls, there
will be length scales which are larger than $vt_b$ (at percolation,
there will be domains of size $H^{-1}$). 
However, the spatial correlation
within these domains falls rapidly beyond a distance of the order
of the typical bubble radius \cite{mm20}, so we do not expect a relevant length scale beyond this distance. 
Moreover, for the single-bubble contribution, any length scale involved is of order $ vt_b $ or smaller, so
the time scale is the relevant quantity.
Indeed, this contribution alone has a peak at
$\omega\sim t_b^{-1}$ rather than at $ \omega\sim d_b^{-1} $ (see Fig.~\ref{figpartes} in Ap.~\ref{apsimult} for the simultaneous case or Ref.~\cite{jt17} for the exponential case).

\subsection{Model comparison}

Since two different nucleation rates depend on different kinds of parameters, 
for a sensible comparison it
is necessary to fix some physical quantity. 
For the exponential and delta-function cases, the
final average bubble separation $d_{b}\equiv n_{b}^{-1/3}$ has often
been used for such a purpose (see, e.g., \cite{w16,chw18}). For the
simultaneous nucleation, this is just a parameter of the nucleation
rate, while for the exponential nucleation it is given 
by 
\begin{equation}
	d_{b}=(8\pi)^{1/3}v/\beta,
	\label{dbbeta}
\end{equation}
If we fix a different physical quantity,
the comparison will be quantitatively different. 
For instance, we may compare an exponential nucleation and a simultaneous nucleation
for which the phase transition has the same duration.
We may
take, as an estimation, the time $\Delta t=t_{2}-t_{1}$ between
the moment $t_{1}$ at which $f_{+}=0.99$
and $t_{2}$ at which $f_+=0.01$ \cite{mm20}. We have $I(t_{1})=-\log(0.99)\equiv I_{1}$
and $I(t_{2})=-\log(0.01)\equiv I_{2}$. For the exponential nucleation
rate this quantity is given by
\begin{equation}
\Delta t=(\log I_{2}-\log I_{1})\beta^{-1}\simeq6.13\beta^{-1},\label{deltatexp}
\end{equation}
while for the simultaneous nucleation it is given by
\begin{equation}
\Delta t=(3/4\pi)^{1/3}(I_{2}^{1/3}-I_{1}^{1/3})(d_{b}/v)\simeq0.898(d_{b}/v).\label{deltatsim}
\end{equation}
For the same  $\Delta t$, the relation between the parameters of these models is
\begin{equation}
d_{b}=\left(\frac{4\pi}{3}\right)^{1/3}\frac{\log I_{2}-\log I_{1}}{I_{2}^{1/3}-I_{1}^{1/3}}\frac{v}{\beta}\label{dbbetadt}
\end{equation}
instead of (\ref{dbbeta}).

\begin{figure}[tb]
	\centering
	\includegraphics[width=0.6\textwidth]{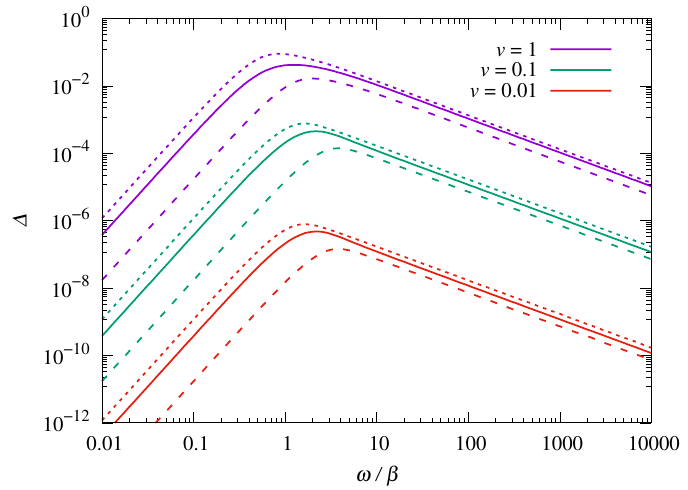}
	\caption{The GW spectrum for an exponential nucleation rate (solid lines), a delta-function nucleation rate with parameter
		$d_{b}$ given by Eq.~(\ref{dbbeta}) (dashed lines), and a delta-function
		rate with $d_{b}$ given by Eq.~(\ref{dbbetadt}) (dotted lines).
		\label{figspectra}}
\end{figure}

In Fig.~\ref{figspectra} we compare the two models fixing either $ d_b $ or $ \Delta t $. 
We need to use the same unit of frequency for all the curves, and we chose $\omega_{*}=\beta$. 
The dimensionless spectrum $\Delta$ is also normalized using $\omega_{*}=\beta$ in Eq.~(\ref{Deltadef}),
which thus coincides with Eq.~(\ref{Deltadef-exp}). 
We consider an exponential nucleation rate
(solid lines), a simultaneous nucleation with $d_{b}$ given by Eq.~(\ref{dbbeta})
(dashed lines), and a simultaneous nucleation with $d_{b}$ given
by Eq.~(\ref{dbbetadt}) (dotted lines), for three different values of the wall velocity. 
The fact that there is not a unique way of comparing two different
nucleation rates implies that the position of the peak for the simultaneous
case can be either to the left or to the right of the peak for the
exponential case, depending on the quantity which is fixed in the
comparison. Similarly, the peak amplitude can be higher or lower.

For a given wall velocity, we see that the GW spectrum for an exponential nucleation
is quite closer to that for a simultaneous nucleation with the same value of
$\Delta t$ than to one with the same value of $d_b$. 
This seems to be another indication of the fact that, for the bubble collision mechanism,
the time scale is more relevant than the length scale. 
Furthermore, 
we verify that the peak frequency $\omega_{p}$ is 
within the range $(1-3)\beta$ for all the curves in Fig.~\ref{figspectra},
while  $d_{b}\sim v\beta^{-1}$
varies by two orders of magnitude for the velocity range considered in the figure.
On the other hand, the amplitude $ \Delta_p $ does vary with the distance scale. 
According to Fig.~\ref{figspectra}, we have, roughly, $\Delta_p\sim d_{b}^{3}$.

This behavior
can be seen analytically for an arbitrary nucleation
rate from the approximation $\omega_{p}\simeq\sqrt{3}\omega_{\times}$,
where $ \omega_\times $ is proportional to $ \omega_b=t_b^{-1} $.
According to 
Eqs.~(\ref{wxvchica})-(\ref{wxv1}), as a function of the velocity, the ratio $\omega_\times/\omega_b$
varies between two $\mathcal{O}(1)$
values
(below we discuss the Gaussian case, where the dimensionless coefficients in these equations
depend on the ratio between two time scales). 
Hence, the behavior $ \omega_p\sim\omega_\times\sim t_b^{-1} $ is quite model independent.
For the peak amplitude, we have  $\Delta_{p}\sim \Delta_\times$,
which is roughly $\propto v^{3}$.

\subsection{Two time scales}

The Gaussian nucleation rate $ \Gamma=\Gamma_m e^{-\gamma^2(t-t_m)^2} $ provides a model with two different time scales, 
since the time during which bubble nucleation is active does not necessarily coincide with the duration of the phase transition.
We have defined two time parameters,
namely, the width of the Gaussian, $ t_\Gamma \equiv \gamma^{-1}$,
and the time associated to the minimal bubble separation $ t_{\min} =d_{\min}/v$.
The latter is obtained from $ \Gamma(t) $ alone, i.e., 
omitting the suppression factor $ f_+ $ in Eq.~(\ref{nb}), and is given by
$d_{\min}=(\sqrt{\pi}\Gamma_{m}/\gamma)^{-1/3}$.
We remark that the use of the parameters $ t_\Gamma,t_{\min} $ is convenient due to their simple analytical relations
with the parameters $ \gamma $ and $ \Gamma_m $.
As we discuss in App.~\ref{apgauss}, in the cases in which the phase transition actually occurs around the maximum of the Gaussian, 
$ t_\Gamma $ is directly related to the duration of bubble nucleation and
$ t_{\min} $ is related to the total duration of the phase transition.
Otherwise, the nucleation rate can be approximated by an exponential and there is a single time scale which is not simply related to these parameters.

\begin{figure}[tb]
	\centering
	\includegraphics[width=.33\textwidth]{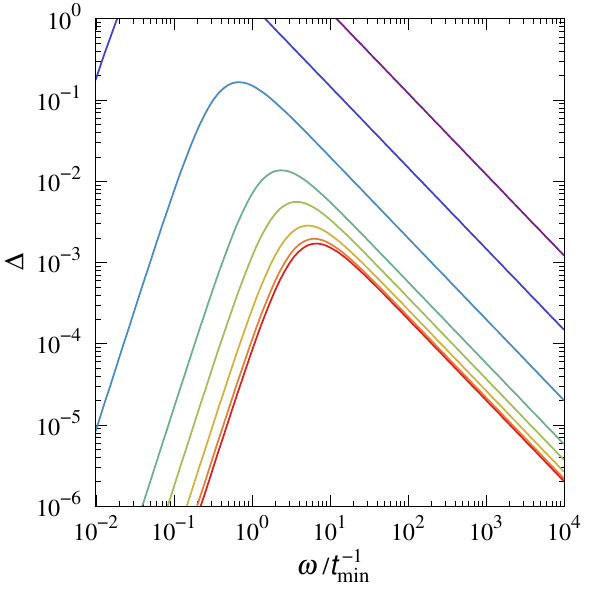}%
	\includegraphics[width=.33\textwidth]{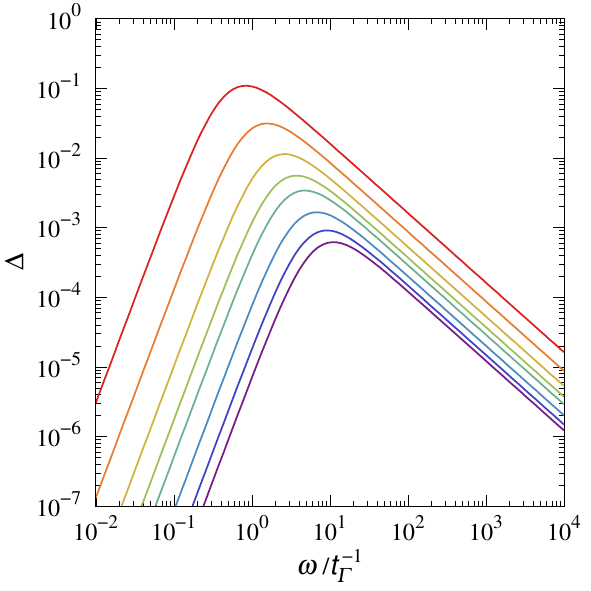}%
	\includegraphics[width=.33\textwidth]{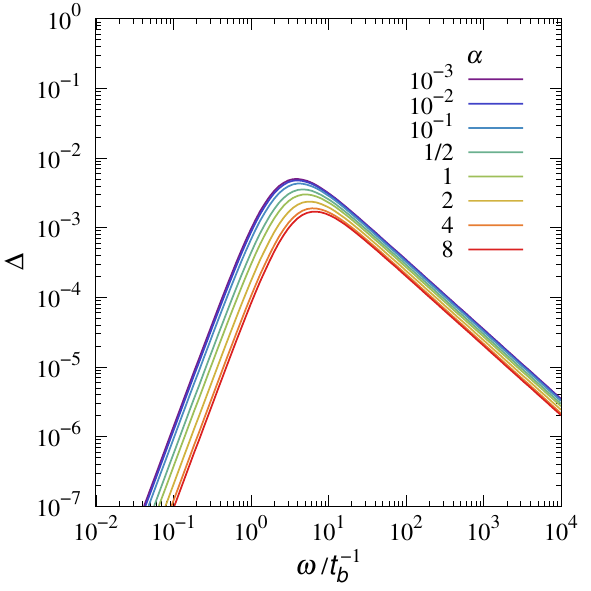}	
	\caption{The spectrum (using the interpolation approximation) for the Gaussian nucleation rate, for  $ v=1 $.
		In the left panel we fix the parameter $ t_{\min} $, in the central panel we fix the parameter $ t_\Gamma $, and in the right panel we fix the parameter $ t_b $. 
		\label{figdeltagaussalfag}}
\end{figure}

In Sec.~\ref{GWgau} we used $ \omega_* =t_{\min}^{-1}$  as the unit of frequency and we considered
different values of the ratio $ \alpha =t_{\min}/t_\Gamma$. 
This means that the different curves in Figs.~\ref{figdeltagauss} and \ref{figdeltagaussalfa} correspond to 
different phase transitions with the same value of $ t_{\min} $. 
In the left panel of Fig.~\ref{figdeltagaussalfag} we consider again the curves of Fig.~\ref{figdeltagaussalfa}.
Here we use the interpolation approximation, so that it is easier to reach wider ranges for $ \omega $ and $ \alpha $.
The central panel shows the same spectra with $ t_\Gamma^{-1} $ as the unit of frequency.
Therefore, in these curves this time scale is fixed while the parameter $ t_{\min} $ varies.
We see that the order of the curves is inverted with respect to those in the left panel.
Like in Fig.~\ref{figspectra}, this shows that, when we compare different models (in this case, different values of $ \alpha $), the result depends on which physical quantity we fix in the comparison.
In the third panel of Fig.~\ref{figdeltagaussalfag} we fix the parameter $ t_b=d_b/v $, 
where $ d_b $ is the real bubble separation, i.e., taking into account the factor $ f_+ $ in Eq.~(\ref{nb}).
Therefore, $ t_b $ is related to the actual duration of the phase transition.
We see that the peak frequency changes very little in this case (we have $ 4\lesssim \omega_p/\omega_*\lesssim 6$ for the whole range of $ \alpha $), indicating that $ \omega_p $ is mostly determined by this time scale. 

\begin{figure}[tb]
	\centering
	\includegraphics[width=.7\textwidth]{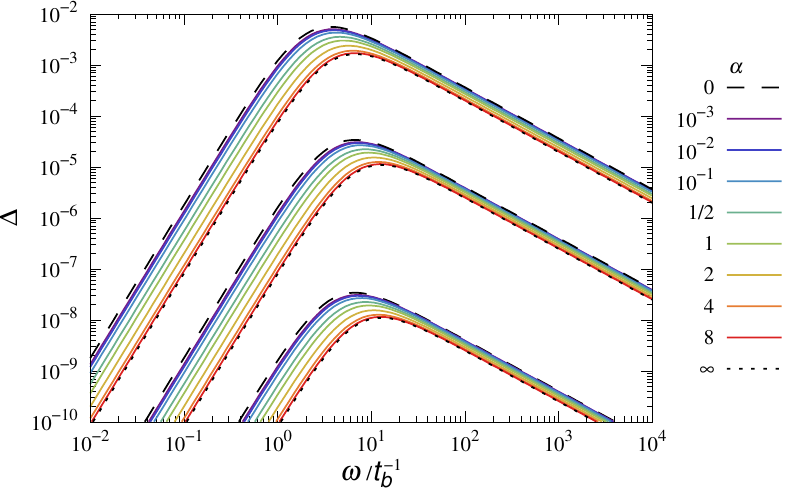}
	\caption{The spectrum (using the interpolation approximation) for the Gaussian nucleation rate, for different values of $ \alpha $ and $ v $. The reference frequency is $ \omega_*=t_b^{-1} $.  The three groups of curves correspond to velocities $ v=1$, $0.1$, and $0.01 $ from top to bottom. 
		The dotted lines indicate the limit $\alpha\to\infty $ (simultaneous nucleation), and the dashed lines indicate the limit $ \alpha\to 0 $ (exponential nucleation).
		\label{figdeltagaussalfav}}
\end{figure}

As already mentioned, in the limit $ \alpha\to\infty $ the width of the Gaussian becomes very small in comparison with the duration of the phase transition
($ t_\Gamma\ll t_{\min} <t_b$), and we have a simultaneous nucleation. Also, we have $ t_{\min}\to t_b $ in this limit. 
The convergence to the simultaneous case can be seen in the left and right panels of Fig.~\ref{figdeltagaussalfag}, but not in the central panel, where we normalize the frequency and amplitude using $ t_\Gamma^{-1} $.
On the other hand, in the right panel we also observe a limiting curve for $ \alpha\to 0 $.
In this limit we have $ t_{\min} \ll t_\Gamma$, but in this case  none of these parameters correspond to a physical time in the evolution of the phase transition.
This limit is obtained for $ \gamma\to 0 $ or $ \Gamma_m\to\infty $.
Since the nucleation rate acts from $ t=-\infty $, this means that in this case the phase transition will be completed 
about a time $ t_* $  much earlier than $ t_m $.
Around the time $ t_* $ the usual exponential approximation can be used,
and the limiting curve corresponds to an exponential nucleation rate  (see Ap.~\ref{apgauss} for more details).
Since the value of $ d_b $ is fixed in the right panel of Fig.~\ref{figdeltagaussalfag},  
the parameter of the exponential is given by $ \beta=(8\pi)^{1/3}v/	d_{b} $.
In Fig.~\ref{figdeltagaussalfav} we indicate this limiting curve with a dashed line, and the limit for $ \alpha\to\infty $
with a dotted line.
We also consider different values of the wall velocity, and we observe again that the peak frequency is always determined by the time scale $ t_b $, while the peak amplitude depends on the length scale $ d_b=v t_b $.	

\subsection{Relation with the surface area}

Since the characteristic frequency of the gravitational waves is determined by the time scale of the source, it is useful to consider the time evolution of the latter.
For the envelope approximation it is clear that the source of GWs is the motion of thin walls. 
More generally, for any mechanism associated to the bubble walls, 
the GW production will be weighted by the amount of bubble wall which
is present at a given time (as can be seen from the general expressions derived in Ref.~\cite{mm21a}).
One may wonder
whether the relevant quantity here is the average uncollided wall area, $ \langle S\rangle\sim R^2 f_+\sim R^2 e^{-4\pi R^3/3} $,
or the surface energy $ \sigma \langle S\rangle\sim R^3 e^{-4\pi R^3/3} $ 
(since $ \sigma\propto R $ due to the release of latent heat),
or some other quantity.
In the envelope approximation, the energy-momentum tensor is given by Eq.~(\ref{Tijenv}),
$ T_{ij}=\sigma\delta(r-R)\,\hat{r}_{i}\hat{r}_{j}\,1_{S}(\hat{r}) $.
Since $ \langle 1_S \rangle = \langle S \rangle/4\pi R^2$, 
this seems to indicate that
the ``effective'' surface energy density
$ \sigma \langle S \rangle/R^2 \sim R e^{-4\pi R^3/3} $ is the relevant quantity.
Let us consider for simplicity the case of simultaneous nucleation, for which the 
above expressions are exact since all the bubbles have the same radius. For instance, the
average uncollided wall area of a bubble is just given by 
\begin{equation}
	\langle S\rangle=4\pi R^2e^{-\frac{4\pi}{3} R^3} =4\pi d_{b}^{2}\left(\frac{t-t_{*}}{t_{b}}\right)^{2}\exp\left[-\frac{4\pi}{3}\left(\frac{t-t_{*}}{t_{b}}\right)^{3}\right],\label{Smedia}
\end{equation}
and the total uncollided wall area per unit volume is given by
$n_{b}\langle S\rangle=d_{b}^{-3}\langle S\rangle$.
We observe a qualitative relation between this quantity and the
GW spectrum, namely, that the time variation is determined by $ t_b $, while the amplitude is determined
by the characteristic $ d_b $. 
Notice, however, that the GW spectrum involves the correlator $ \langle T_{ij}T_{kl}' \rangle $. 
For a given bubble, any of the quantities discussed above is of the form $ R^n S $, and 
we should expect that the result depends on  $ R^n R^{\prime n} \langle SS' \rangle$, i.e., 
that the relevant quantity is the surface correlation rather than $ \langle S \rangle $.

In the expression for $ T_{ij} $, the time independent factor $ \hat r_i\hat r_j $  characterizes the spatial dependence of the source (the spherical wall),
while the time dependence is contained essentially in the variation of the bubble radius $ R$ and the uncollided surface $S $.
If we omit the factor $\Lambda_{ij,kl} \hat r_i\hat r_j \hat r_k'\hat r_l' $ in Eq.~(\ref{defPi}),  
and consider for simplicity only the single-bubble contribution, we obtain the rough estimation
\begin{equation}
	\Pi(t,t',\omega)\sim n_b\,\sigma\sigma' \left\langle S(t)S(t')\right\rangle 
	\sim n_{b}\left(\kappa\rho_{\mathrm{vac}}\right)^{2}RR'\left\langle S(t)S(t')\right\rangle .
	\label{PiSS}
\end{equation}
In Ref.~\cite{mm20} it was argued that
an estimation for the GW spectrum can be obtained by assuming
that the quantity $\Pi(t,t',\omega)$ is proportional to the surface
correlator $\left\langle S(t)S(t')\right\rangle $. 
This is equivalent to the further approximation $ R\sim R'\sim d_b $ in Eq.~(\ref{PiSS}).
For the simultaneous case and the single bubble contribution we have
\cite{mm20} 
\begin{equation}
\langle S(t)S(t')\rangle=8\pi^{2}RR^{\prime}\int_{R'-R}^{R'+R}ds\,s\,e^{-I_{\mathrm{tot}}(t,t',s)}.\label{SSp}
\end{equation}
(notice that this quantity contains information about the 
correlation between points on the bubble surface).
Proceeding as before, for the approximation (\ref{PiSS}) we obtain 
\begin{equation}
\Delta\sim
v^{3}\tilde{\omega}^{3}\int_{0}^{\infty}d\tau_{+}\int_{0}^{\tau_{+}}d\tau_{s}\,\tau_{s}\,\int_{0}^{\tau_{s}}d\tau_{-}
\cos(\tilde{\omega}\tau_{-}) \left(\tau_{+}^{2}-\tau_{-}^{2}\right)^2 e^{-I_{\mathrm{tot}}(\tau_{+},\tau_{-},\tau_{s})}.
\label{deltasimultap}
\end{equation}
The integrand does not depend on the velocity, so we have an amplitude
proportional to $v^{3}$ and a shape which depends on $\tilde{\omega}=\omega t_{b}$.
The computation of Eq.~(\ref{deltasimultap}) is much simpler than
the complete expression 
(\ref{Deltassimultfinal}) (see appendix 
\ref{apsimult}). 
We show the result in Fig.~\ref{figdeltaap} for $v=1$ (dashed black
line). The approximation gives the correct power-law form\footnote{The additional approximation $\left\langle S(t)S(t')\right\rangle \sim\left\langle S(t)\right\rangle \left\langle S(t')\right\rangle $
	simplifies significantly the expressions, but does not give the correct
	behavior at high frequencies. It is a smoother function of time, and
	therefore its Fourier transform falls more rapidly at high $\omega$.}. 
The peak intensity, though, is a few orders
of magnitude higher than the exact result (solid line). This is the
effect of ignoring the spatial dependence of the source, since the
spherical shape of the bubbles causes a suppression. In particular,
in the envelope approximation the sphericity is lost at the expense of decreasing
the surface area of the walls which produce the gravitational radiation.
\begin{figure}[tb]
\centering
\includegraphics[width=0.7\textwidth]{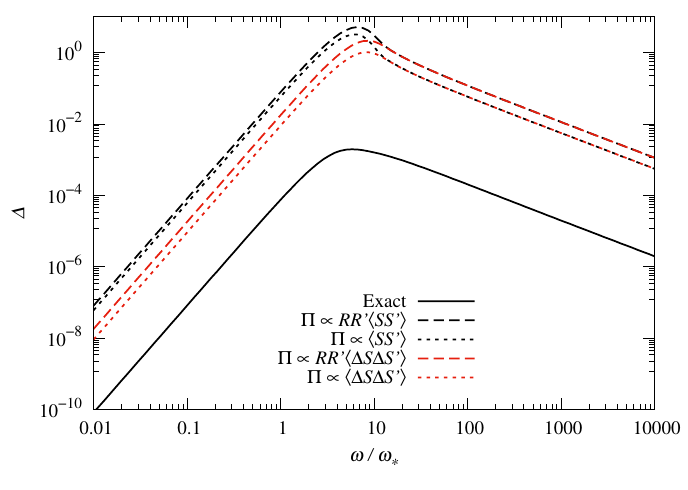}
\caption{The GW spectrum and some rough approximations, for a simultaneous
nucleation and $v=1$ ($ \omega_{*}=v/d_b $).
\label{figdeltaap}}
\end{figure}

If we replace $ RR' $ by $ d_b^2 $ in Eq.~(\ref{PiSS}),  we obtain
a single factor of $ (\tau_+^2-\tau_-^2) $ in Eq.~(\ref{deltasimultap}).
This result is plotted with a dotted black line in Fig.~\ref{figdeltaap}.
We see that the spectrum does not change significantly, indicating that the relevant quantity which determines its shape 
is $ \langle SS' \rangle $.
In \cite{mm20} we argued that it would be more realistic to relate
the GW spectrum to the quantity $\langle\Delta S(t)\Delta S(t')\rangle$,
with $\Delta S=S-\langle S\rangle$, rather than to $\langle S(t)S(t')\rangle$,
in order to take into account the fact that the result should vanish
if the surfaces at $t$ and $t'$ were uncorrelated. Indeed, the separation
$\left\langle S(t)S(t')\right\rangle =\left\langle S(t)\right\rangle \left\langle S(t')\right\rangle $
corresponds to assuming that any two points on the bubble surface
are not correlated, i.e., to approximating the probability 
$e^{-I_{\mathrm{tot}}}=e^{-I(t)}e^{-I(t')}e^{I_{\cap}(t,t',s)}$
by $e^{-I(t)}e^{-I(t')}$. 
Replacing  $S$ with $\Delta S$ in Eq.~(\ref{PiSS}) we obtain
the red lines in Fig.~\ref{figdeltaap} (with and without taking into account the extra factor of $ RR' $).
Only the
low-frequency part of the spectrum changes with respect to the previous approximations. 
The shape of the spectrum
is more realistic, but it is quantitatively very similar.

\section{Conclusions and discussion}

\label{conclu}

We have studied the general features of gravitational waves from bubble collisions in the  envelope approximation.
For that aim we have applied the approach of Ref.~\cite{mm21a} to this particular mechanism of GW generation.
In the first place, 
we have computed the GW spectrum for several phase transition models.
Our results for these specific models are in agreement with previous works, whereas 
our analytic expressions allowed us to consider wider frequency ranges
as well as grater variations of parameters.

In second place, 
we have studied the asymptotic limits of the 
spectrum for arbitrary nucleation rate $ \Gamma(t) $ and wall velocity $ v(t) $.
We have thus confirmed analytically that the GW spectrum for the envelope approximation
always rises as $\omega^{3}$ for low frequencies and falls as $\omega^{-1}$
for high frequencies, independently of the specific evolution of the
phase transition. 
Therefore, in the two ends of the spectrum, it is only necessary to compute numerically the constant coefficients of
these power laws.
For constant velocity, the calculation of these coefficients simplifies significantly,
and we obtained analytically the dependence on $ v $.
Furthermore, we provided a simple interpolation between
the asymptotes, which can be used as an estimate of the whole spectrum, thus
avoiding difficult numerical computations.
These analytic approximations are useful for studying the dependence of the spectrum
on the parameters of the model. 
Although in this work we have focused on the envelope approximation,
we expect that our determination of the asymptotes can be generalized to other mechanisms such 
as the bulk flow model \cite{jt19,k18}, where it is particularly difficult to compute the spectrum 
at high frequencies.

Finally, we have used our results to study the dependence of the GW spectrum on the characteristic time and distance scales
of the phase transition.
We have confirmed that the peak frequency $ \omega_p $ is generally determined by the time scale rather than the length scale.
More precisely, we have $ \omega_p\sim t_b^{-1} $, where $ t_b $ is the \emph{total duration} of the phase transition.
We have checked this fact, both numerically and analytically,  by varying the bubble size $ d_b=vt_b $ as well as the time 
$ t_\Gamma \leq t_b $ during which bubble nucleation is active.
The amplitude of the spectrum does depend on the size scale $ d_b $ (the dimensionless spectrum $ \Delta $ goes roughly as $ v^3 $).
We have related these features to the time correlation of the uncollided wall surface area, which is essentially the source of GWs
in the envelope approximation.
Moreover, the rough approximation $\Pi\propto\langle S(t) S(t')\rangle$,
which corresponds essentially to neglecting the spatial dependence
of the source, gives the correct position of the peak as well as the
correct behavior at low and high frequencies, although the amplitude
is a few orders of magnitude too high.

Lattice simulations of vacuum bubbles generally give a different form
of the spectrum. In the first place, in Ref.~\cite{cg12}, it was
found that GWs are produced after bubble percolation. However, in
Ref.~\cite{chw18} this effect was identified with oscillations of
the scalar field, which produce GWs with a frequency of the order
of the scalar mass in the broken phase. Thus, with a realistic separation
of scales (not achievable in the simulation), this signal will be
actually at a much higher frequency, and will have a much smaller amplitude.

In Ref.~\cite{w16} it was found that the GW spectrum sourced by
the scalar field agrees in shape and intensity with the envelope approximation,
at least in the frequency range delimited by the inverses of the box
size $L$ of the simulation and the wall width $l_{w}$. At higher
frequencies the decrease becomes steeper. In the more recent simulation
\cite{chw18}, it was found that the peak of the spectrum from bubble
collisions is slightly shifted towards the infrared with respect to
the envelope approximation. Besides, the power law on the high frequency
side of this peak seems to be given by $b\simeq1.5$, in contrast
to the value $b=1$ for the envelope approximation. In the subsequent
work \cite{cghw20}, the wall thickness was varied (by changing a
parameter in the effective potential), and it was found that for thicker
walls the ultraviolet power law has even larger values, up to $b\simeq2.3$.
The more recent work \cite{gsw21} supports the conclusion that the high-frequency power law becomes steeper
for thick-walled bubbles.

However, we remark that the bubble radius and the wall width, which
in general differ by several orders of magnitude, in the lattice simulations
are separated by, at most, a couple of orders of magnitude. For instance,
in \cite{chw18}, the power law $b\simeq1.5$ fits the curve in a
frequency range of one order of magnitude between the peak frequency
$\omega_{p}$ and $\omega\sim10\omega_{p}$. A little beyond this
range is the ultraviolet bump due to the field oscillations. This
second peak is associated to the scalar mass scale, which is of the
same order of the wall width $l_{w}$, so this part of the curve is
also influenced by this parameter. Therefore, it is possible that
the differences with the envelope approximation at high frequency
are due to an insufficient separation of these scales.

In order to avoid this problem, in Refs.~\cite{lv20b,lv21} a two-bubble collision is first studied by lattice simulations
to determine how the surface energy density  scales with the bubble radius in the collided regions, and then
the GW spectrum is computed in many-bubble thin-wall simulations like those of Ref.~\cite{k18} for the bulk flow model.
The power law exponents obtained with this approach are also different from the envelope approximation and depend on the 
nature of the scalar field, i.e., they are different for a real scalar field, a complex scalar field which breaks a $ U(1) $ global symmetry, and the case of a gauge $ U(1) $ symmetry. 
As already mentioned, the part of this calculation which is equivalent to the bulk flow model 
can be approached semi-analytically, but the numerical integrals become very difficult for frequencies higher than the peak \cite{jt19}.
Therefore, the high-frequency approximations we used to obtain the asymptotic behavior for the envelope case 
will be useful to address this kind of calculation.

\section*{Acknowledgments}

This work was supported by CONICET grant PIP 11220130100172 
and Universidad Nacional de Mar del Plata, grant EXA999/20.

\appendix

\section{Details for the simultaneous case}

\label{apsimult}

In this appendix we present some analytic and numerical results for
the delta-function nucleation rate.

\subsection{Gaussian integrals}

The integrals 
\begin{equation}
\tilde{Q}(\tau_{+},\tau_{s},\tilde{\omega})=\int_{0}^{\tau_{s}}d\tau_{-}\cos(\tilde{\omega}\tau_{-})Q(\tau_{+},\tau_{-},\tau_{s})e^{-\frac{\pi}{4}\frac{(\tau_{+}+\tau_{s})^{2}}{\tau_{s}}\tau_{-}^{2}}\label{Qapen}
\end{equation}
 and 
\begin{equation}
\tilde{P}_{i}(\tau_{+},\tau_{s},\tilde{\omega})=\int_{0}^{\tau_{s}}d\tau_{-}\cos(\tilde{\omega}\tau_{-})P_{i}(\tau_{+},\tau_{-},\tau_{s})e^{-\frac{\pi}{4}\frac{(\tau_{+}+\tau_{s})^{2}}{\tau_{s}}\tau_{-}^{2}},\label{Piapen}
\end{equation}
where $Q$ and $P_{i}$ are the polynomials defined in Eqs.~(\ref{Q})
and (\ref{P0}-\ref{P2}), are straightforward, since the polynomials
are of the form $A+B\tau_{-}^{2}+C\tau_{-}^{4}+D\tau_{-}^{6}$ (with
$D=0$ for the $P_{i}$) and the cosine can be written as a combination
of exponentials. We obtain 
\begin{align}
\tilde{Q}= & \,(\tau_{+}^{2}-\tau_{s}^{2})^{2}\left[\tau_{s}^{8}I^{(0)}-\tau_{s}^{4}\left(\tau_{+}^{2}+2\tau_{s}^{2}\right)I^{(2)}+\tau_{s}^{2}\left(2\tau_{+}^{2}+\tau_{s}^{2}\right)I^{(4)}-\tau_{+}^{2}I^{(6)}\right],\nonumber \\
\tilde{P}_{0}= & \,(\tau_{+}^{2}-\tau_{s}^{2})^{2}\left(\tau_{s}^{4}I^{(0)}-2\tau_{s}^{2}I^{(2)}+I^{(4)}\right),\\
\tilde{P}_{1}= & \,2(\tau_{+}^{2}-\tau_{s}^{2})\left[-\left(\tau_{+}^{2}+3\tau_{s}^{2}\right)\tau_{s}^{4}I^{(0)}+2\left(3\tau_{+}^{2}+\tau_{s}^{2}\right)\tau_{s}^{2}I^{(2)}+\left(\tau_{s}^{2}-5\tau_{+}^{2}\right)I^{(4)}\right],\nonumber \\
\tilde{P}_{2}= & \,(3\tau_{s}^{4}+2\tau_{+}^{2}\tau_{s}^{2}+3\tau_{+}^{4})\tau_{s}^{4}I^{(0)}+2(\tau_{s}^{4}+6\tau_{+}^{2}\tau_{s}^{2}-15\tau_{+}^{4})\tau_{s}^{2}I^{(2)}+(3\tau_{s}^{4}-30\tau_{+}^{2}\tau_{s}^{2}+35\tau_{+}^{4})I^{(4)}.\nonumber 
\end{align}
where the quantities $I^{(n)}$ are the integrals 
\begin{equation}
I^{(n)}=\int_{0}^{\tau_{s}}d\tau_{-}\,\tau_{-}^{n}\cos(\tilde{\omega}\tau_{-})e^{-\alpha\tau_{-}^{2}}=\mathrm{Re}\left[\int_{0}^{\tau_{s}}d\tau_{-}\,\tau_{-}^{n}e^{-\alpha\tau_{-}^{2}-i\tilde{\omega}\tau_{-}}\right],\label{integgaus}
\end{equation}
with $\alpha=\frac{\pi}{4}(\tau_{+}+\tau_{s})^{2}/\tau_{s}$. These
are given by 
\begin{equation}
I^{(0)}=\sqrt{\frac{\pi}{4\alpha}}\,\exp\left(-\frac{\tilde{\omega}^{2}}{4\alpha}\right)\,\mathrm{Re}\left[\mathrm{erf}\left(\sqrt{\alpha}\tau_{s}+\frac{i\tilde{\omega}}{2\sqrt{\alpha}}\right)\right],
\end{equation}
\begin{equation}
I^{(2)}=\frac{1}{4\alpha^{2}}\left[\left(2\alpha-\tilde{\omega}^{2}\right)I^{(0)}+S-C\right],
\end{equation}
\begin{align}
I^{(4)}= & \,\frac{1}{16\alpha^{4}}\left[\left(\tilde{\omega}^{4}-12\alpha\tilde{\omega}^{2}+12\alpha^{2}\right)I^{(0)}\right.\nonumber \\
 & \left.-\left(\tilde{\omega}^{2}-4\alpha^{2}\tau_{s}^{2}-10\alpha\right)S+\left(\tilde{\omega}^{2}-4\alpha^{2}\tau_{s}^{2}-6\alpha\right)C\right],
\end{align}
\begin{align}
I^{(6)}= & \,\frac{1}{64\alpha^{6}}\left[\left(-\tilde{\omega}^{6}+30\alpha\tilde{\omega}^{4}-180\alpha^{2}\tilde{\omega}^{2}+120\alpha^{3}\right)I^{(0)}\right.\nonumber \\
 & +\left(\tilde{\omega}^{4}-4\alpha^{2}\tau_{s}^{2}\tilde{\omega}^{2}-28\alpha\tilde{\omega}^{2}+16\alpha^{4}\tau_{s}^{4}+72\alpha^{3}\tau_{s}^{2}+132\alpha^{2}\right)S\nonumber \\
 & \left.-\left(\tilde{\omega}^{4}-4\alpha^{2}\tau_{s}^{2}\tilde{\omega}^{2}-24\alpha\tilde{\omega}^{2}+16\alpha^{4}\tau_{s}^{4}+40\alpha^{3}\tau_{s}^{2}+60\alpha^{2}\right)C\right],
\end{align}
where 
\begin{equation}
S=e^{-\alpha\tau_{s}^{2}}\tilde{\omega}\sin(\tilde{\omega}\tau_{s}),\;C=2\alpha\tau_{s}e^{-\alpha\tau_{s}^{2}}\cos(\tilde{\omega}\tau_{s}),
\end{equation}
and $\mathrm{Re}[\mathrm{erf}(z)]$ is the real part of the error
function.

On the other hand, for the approximation (\ref{deltasimultap}), we
have to calculate the integral 
\begin{equation}
\tilde{P}=\int_{0}^{\tau_{s}}d\tau_{-}\cos(\tilde{\omega}\tau_{-})\left(\tau_{+}^{2}-\tau_{-}^{2}\right)^2
e^{-\frac{\pi}{4}\frac{(\tau_{+}+\tau_{s})^{2}}{\tau_{s}}\tau_{-}^{2}}.
\label{Ptil}
\end{equation}
which has the same form of Eq.~(\ref{Piapen}), but for the function
$P=\tau_{+}^{4}-2\tau_+^2\tau_{-}^{2} +\tau_-^4$ instead of $P_{i}$.
We obtain in this case $\tilde{P}=\tau_+^{4}I^{(0)}-2\tau_+^{2}I^{(2)}+I^{(4)}$, and Eq.~(\ref{deltasimultap}) is given by
\begin{equation}
\Delta\sim v^{3}\tilde{\omega}^{3}\int_{0}^{\infty}d\tau_{+}\int_{0}^{\tau_{+}}d\tau_{s}\,\tau_{s}\,\tilde{P}\,e^{-\frac{\pi}{12}\left(2\tau_{+}^{3}-\tau_{s}^{3}+3\tau_{+}^{2}\tau_{s}\right)}.\label{deltasimultapapend}
\end{equation}
If $ RR' $  is replaced by $ d_b^2 $ in Eq.~(\ref{PiSS}),  then we have 
a single factor of $ (\tau_+^2-\tau_-^2) $ in Eq.~(\ref{deltasimultap}),
so we have to calculate the integral (\ref{Ptil}) with
$P=\tau_{+}^{2}-\tau_{-}^{2}$, and
we obtain in this case $\tilde{P}=\tau_{+}^{2}I^{(0)}-I^{(2)}$.
If $\langle\Delta S(t)\Delta S(t')\rangle$ is used instead of $\langle S(t)S(t')\rangle$
in Eq.~(\ref{PiSS}), then, in the integrand of Eq.~(\ref{deltasimultapapend}),
we have to subtract, to the quantity $\tilde{P}\,e^{-I_{\mathrm{tot}}}$,
the same quantity evaluated at $\tau_{s}=\tau_{+}$.

\subsection{Contributions to the spectrum}

\begin{figure}[tbh]
\centering
\includegraphics[width=0.7\textwidth]{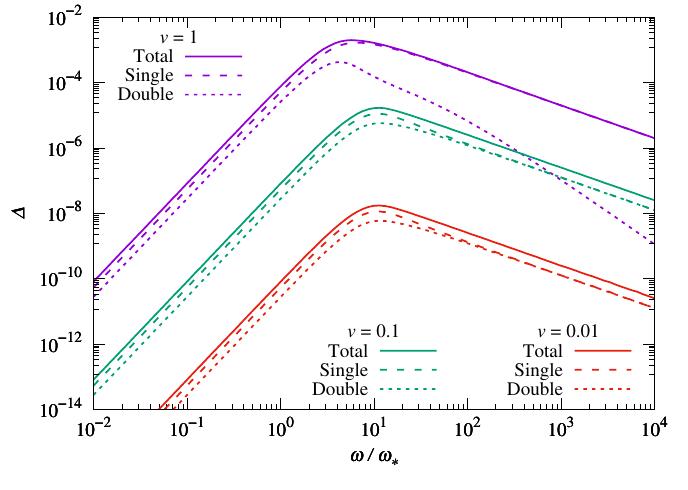}
\caption{The different contributions to the spectrum for the case of simultaneous nucleation,
for a few values of the wall velocity $v$ ($\omega_{*}=v/d_{b}$).
\label{figpartes}}
\end{figure}

It is of interest to show separately the single-bubble and the two-bubble
contributions. In Fig.~\ref{figpartes} we consider the case of simultaneous
nucleation for a few velocities (the exponential case was already
considered in Ref.~\cite{jt17}). We see that the two contributions
are in general of the same order, except for the case $v=1$, where
the intensity falls as $\omega^{-2}$ for $\omega/\omega_{b}\gg1$,
as shown analytically in Sec.~\ref{bbcoll}.

\section{Details for the Gaussian case}
\label{apgauss}

In Sec.~\ref{GWgau} we have considered as a reference time the parameter $ t_{\min} $ and as a reference frequency 
$\omega_* = t_{\min}^{-1} $. 
Thus, in Eqs.~(\ref{Dsvconst})-(\ref{G}), we have 
$ \tilde \omega \equiv \omega / \omega_* = \omega t_{\min}$, and the dimensionless times are given by Eq.~(\ref{cambiovar}) with $ t_*=t_m $ and $ t_b=t_{\min} $. For instance, we have  $ \tau=(t-t_m)/t_{\min} $.
The dimensionless rate is given by $ \tilde \Gamma(\tau)=({\alpha}/{\sqrt{\pi}}) e^{-(\alpha\tau)^2} $, and the integrals in Eqs.~(\ref{F}) and (\ref{G}) 
can be done analytically, as well as those in Eqs.~(\ref{It}) and (\ref{Itts}). 
To simplify a little the expressions, in this appendix we shall consider the parameter $ t_\Gamma $ 
as the reference time, and the corresponding reference frequency $ \omega_* =t_\Gamma^{-1}=\gamma$.
To avoid confusion, we shall denote the  dimensionless spectrum (\ref{Deltadef}) as $ \bar\Delta $, and we shall denote $ \bar\omega \equiv \omega/\omega_* = \omega t_\Gamma $.
The relation with $ \tilde \omega $ and the corresponding spectrum 
is $ \bar\omega=\tilde\omega /\alpha$,
$ \bar\Delta(\bar\omega)= \alpha^2 \Delta (\alpha\bar\omega)$.
We also denote the corresponding dimensionless times by $ x=(t-t_m) /t_\Gamma$, $ x_N=(t_N-t_m)/t_\Gamma $, etc.,
which corresponds to the change of variables $ x=\alpha\tau , x_N=\alpha\tau_N $, etc.\ in 
Eqs.~(\ref{Dsvconst})-(\ref{G}).

\subsection{Formulas for the spectrum and the asymptotes}

We obtain
\begin{equation}
	\bar\Delta^{(s)}=\frac{v^{3}\bar{\omega}^{3}}{48\alpha^3}
	\int_{-\infty}^{\infty}dx_{+}\int_{0}^{\infty}dx_{-}\cos({\bar{\omega}x_{-}})
	\int_{x_{-}}^{\infty} \frac{dx_{s}}{x_{s}^{3}}
	\sum_{i=0}^{2}\frac{j_{i}(v\bar{\omega}x_{s})}{(v\bar{\omega}x_{s})^{i}}F_{i}
	e^{-I_{\mathrm{tot}}}.
\end{equation}
\begin{equation}
	\bar\Delta^{(d)}=\frac{\pi v^{3}\bar{\omega}^{3}}{48\alpha^{6}}
	\int_{-\infty}^{\infty}dx_{+}\int_{0}^{\infty}dx_{-} \cos(\bar{\omega}x_{-})
	\int_{x_{-}}^{\infty}\frac{dx_{s}}{x_{s}^{4}}
	\frac{j_{2}(v\bar{\omega}x_{s})}{(v\bar{\omega}x_{s})^{2}}G_{+}G_{-}e^{-I_{\mathrm{tot}}},
\end{equation}
where the functions $ F_i $, $ G_\pm $, and $ I_\mathrm{tot} $ are given by 
\begin{align}
	F_{0}= &	\,\frac{(x_{-}^{2}-x_{s}^{2})^{2}}{2} 
	\left\{ \left(x_{s}^{4}+x_{+}^{4}-4x_{s}^{2}-2x_{+}^{2}x_{s}^{2}+12x_{+}^{2}+12\right) 
	\left[\text{erf} \left( \frac{x_{+}-x_{s}}{2}\right)+1\right]\right. \nonumber
	\\
	& \, \left. +
	\left(x_{+}^{3}-x_{s}^{3}+x_{+}^{2}x_{s}-x_{+}x_{s}^{2}+10x_{+}+6x_{s}\right)
	\frac{2}{\sqrt{\pi}}e^{-\frac{1}{4}\left(x_{+}-x_{s}\right)^{2}}\right\} ,
\end{align}
\begin{align}
	F_{1}= & \,
	\left(x_{s}^{2}-x_{-}^{2}\right)
	\left\{ \vphantom{\frac{2}{\sqrt{\pi}}}
	\left( 3x_{s}^{6}+x_{-}^{2}x_{s}^{4}-2x_{+}^{2}x_{s}^{4}-x_{+}^{4}x_{s}^{2}+5x_{-}^{2}x_{+}^{4}-6x_{-}^{2}x_{+}^{2}x_{s}^{2} \right. 
	\right. \nonumber
	\\
	& \, -4x_{s}^{4}-12x_{-}^{2}x_{s}^{2}-12x_{+}^{2}x_{s}^{2}+60x_{-}^{2}x_{+}^{2}-12x_{s}^{2}+60x_{-}^{2} \left.\right)  
	\left[\text{erf}\left(\frac{x_{+}-x_{s}}{2}\right)+1\right] 
	 \nonumber
	\\
	& \,  
	-\left( 3x_{s}^{5}+3x_{+}x_{s}^{4}+x_{-}^{2}x_{s}^{3}+x_{+}^{2}x_{s}^{3}-5x_{-}^{2}x_{+}^{2}x_{s}-5x_{-}^{2}x_{+}^{3}
	+x_{+}^{3}x_{s}^{2}+x_{-}^{2}x_{+}x_{s}^{2} \right. \nonumber
	\\
	& \, 
	\left.  \left.  +6x_{s}^{3}+10x_{+}x_{s}^{2}-30x_{-}^{2}x_{s}-50x_{-}^{2}x_{+} \right)
	\frac{2}{\sqrt{\pi}} e^{-\frac{1}{4}\left(x_{+}-x_{s}\right)^{2}} \right\} ,
\end{align}
\begin{align}
	F_{2} & = \left(3x_{s}^{8}+2x_{-}^{2}x_{s}^{6}+2x_{+}^{2}x_{s}^{6}+4x_{s}^{6}+3x_{-}^{4}x_{s}^{4}
	+3x_{+}^{4}x_{s}^{4}+24x_{-}^{2}x_{s}^{4}+12x_{-}^{2}x_{+}^{2}x_{s}^{4}
	 +36x_{+}^{2}x_{s}^{4}\right. \nonumber
	\\
	&  +36x_{s}^{4}-60x_{-}^{4}x_{s}^{2}-30x_{-}^{2}x_{+}^{4}x_{s}^{2}-360x_{-}^{2}x_{s}^{2}
	-30x_{-}^{4}x_{+}^{2}x_{s}^{2}
	-360x_{-}^{2}x_{+}^{2}x_{s}^{2}+420x_{-}^{4}  \nonumber
	\\
	& \left. +35x_{-}^{4}x_{+}^{4} +420x_{-}^{4}x_{+}^{2} \right)
	 \frac{1}{2}\left[\text{erf}\left(\frac{x_{+}-x_{s}}{2}\right)+1\right] \nonumber
	\\
	&  +\left(5x_{s}^{7}+5x_{+}x_{s}^{6}-18x_{-}^{2}x_{s}^{5}+3x_{+}^{2}x_{s}^{5}+18x_{s}^{5}
	+3x_{+}^{3}x_{s}^{4}-18x_{-}^{2}x_{+}x_{s}^{4}+30x_{+}x_{s}^{4}+5x_{-}^{4}x_{s}^{3}
	\right. \nonumber
	\\
	&  -180x_{-}^{2}x_{s}^{3}-30x_{-}^{2}x_{+}^{2}x_{s}^{3} -30x_{-}^{2}x_{+}^{3}x_{s}^{2}+5x_{-}^{4}x_{+}x_{s}^{2} -300x_{-}^{2}x_{+}x_{s}^{2}+210x_{-}^{4}x_{s}
	\nonumber
	\\
	& \left. +35x_{-}^{4}x_{+}^{2}x_{s}+35x_{-}^{4}x_{+}^{3}+350x_{-}^{4}x_{+}\right)
	\frac{1}{\sqrt{\pi}}e^{-\frac{1}{4}\left(x_{+}-x_{s}\right)^{2}}	,
\end{align}
\begin{align}
	G_{\pm}= & \,
	\frac{\left(x_{-}^{2}-x_{s}^{2}\right)}{2}\left\{ \left[x_{s}^{4}-x_{+}^{2}x_{s}^{2}-2x_{s}^{2}\pm x_{-}\left(x_{+}^{3}+6x_{+}-x_{+}x_{s}^{2}\right)\right]
	\left[\text{erf}\left(\frac{x_{+}-x_{s}}{2}\right)+1\right]\right. \nonumber
	\\& \,
	\left.-2\left[x_{s}^{3}+x_{+}x_{s}^{2}\mp x_{-}\left(x_{+}x_{s}+x_{+}^{2}+4\right)\right]\frac{1}{\sqrt{\pi}}e^{-\frac{1}{4}\left(x_{+}-x_{s}\right){}^{2}}\right\} ,
\end{align}
and $ I_\mathrm{tot}=I((x_+-x_-)/2,\alpha)+I((x_++x_-)/2,\alpha)+I_\cap $, with
\begin{equation}
	I(x,\alpha) = \frac{\pi}{3\alpha^3} \left\{ x\left(2x^{2}+3\right)\left[\text{erf}(x)+1\right]
	+\left(x^{2}+1\right)\frac{2}{\sqrt{\pi}}e^{-x^{2}}\right\} 
	\label{Ix}
\end{equation}
and
\begin{align}
	I_{\cap}=&\frac{\pi\alpha^{-3}}{24x_{s}}\left\{ \left(3x_{-}^{2}x_{s}+2x_{+}^{2}x_{s}-x_{s}^{3}-x_{+}x_{s}^{2}+8x_{s}-3x_{-}^{2}x_{+}\right)\frac{2}{\sqrt{\pi}}e^{-\frac{1}{4}\left(x_{+}-x_{s}\right){}^{2}}\right. \nonumber
	\\ 
	&
	+\left(x_{s}^{4}-3x_{-}^{2}x_{s}^{2}-3x_{+}^{2}x_{s}^{2}-6x_{s}^{2}+2x_{+}^{3}x_{s}+6x_{-}^{2}x_{+}x_{s}+12x_{+}x_{s}
	-6x_{-}^{2}-3x_{-}^{2}x_{+}^{2}\right) \nonumber
	\\
	& \left.
    \times \left[\text{erf}\left(\frac{x_{+}-x_{s}}{2}\right)+1\right]\right\} .
\end{align}
The coefficient of the low-frequency asymptote (\ref{Deltalow}) is given by ($ \bar D=\alpha^5 D $)
\begin{equation}
	\bar D=\frac{\alpha^{-3}}{48}
	\int_{-\infty}^{\infty}dx_{+}\int_{0}^{\infty}dx_{-}	
	\int_{x_{-}}^{\infty} \frac{dx_{s}}{x_{s}^{3}}
	\left[F_{0}+\frac{F_1}{3}+\frac{F_2}{15}+\frac{\pi G_{+}G_{-}}{15\alpha^3x_s} \right]
	e^{-I_{\mathrm{tot}}}.
\end{equation}
The coefficients of the high-frequency asymptote (\ref{Deltahigh})-(\ref{Av}) are given by ($ \bar C^{(s)}=\alpha C^{(s)} $)
\begin{equation}
	\bar C^{(s)}=\frac{\pi}{72}
	\int_{-\infty}^{\infty}\negmedspace d\bar x
	e^{-\frac{4\pi}{3}\tilde{I}_{3}(\bar x)}\tilde{I}_{4}(\bar x)\tilde{I}_{2}(\bar x),
	\;
	\bar C^{(d)}=\frac{\pi}{18}\int_{-\infty}^{\infty}\negmedspace d\bar x 
	e^{-\frac{4\pi}{3}\tilde{I}_{3}(\bar x)}\tilde{I}_{3}(\bar x)^{2},
\end{equation}
with
\begin{align}
	\tilde{I}_{2}&=\alpha^{-2}\left\{
	\frac{2\bar{x}^{2}+1}{4}\left[\text{erf}\left(\bar{x}\right)+1\right]+\frac{\bar{x}}{2\sqrt{\pi}}e^{-\bar{x}^{2}}
	\right\},
	\\
	\tilde{I}_{3}&= \alpha^{-3}\left\{
	\frac{\bar{x}\left(2\bar{x}^{2}+3\right)}{4}\left[\text{erf}\left(\bar{x}\right)+1\right]+\frac{\bar{x}^{2}+1}{2\sqrt{\pi}}e^{-\bar{x}^{2}}\right\} ,
	\\
	\tilde{I}_{4}&=\alpha^{-4}\left\{
	\frac{4\bar{x}^{4}+12\bar{x}^{2}+3}{8}
	\left[\text{erf}\left(\bar{x}\right)+1\right]+\frac{\bar{x}\left(2\bar{x}^{2}+5\right)}{4\sqrt{\pi}}e^{-\bar{x}^{2}}
	\right\}.
\end{align}

\subsection{Different parametrizations}

The parametrization 
(\ref{gammagaussini}) for the nucleation rate,
\begin{equation}
	\Gamma(t)=\Gamma_{m}e^{-\gamma^{2}(t-t_{m})^{2}},
	\label{gammagaussnormal}
\end{equation} 
is centered at the maximum of the Gaussian. 
We have conveniently defined the time parameters $ t_\Gamma=\gamma^{-1} $ and 
$ t_{\min}= d_{\min}/v =v^{-1}(\gamma/\sqrt{\pi}\Gamma_m)^{1/3}$ 
in order to obtain the simple expression (\ref{gammagauss}) where the dimensionless rate $ \tilde\Gamma $ depends only on the parameter $ \alpha=t_{\min} /t_\Gamma $. 
In terms of the basic parameters of the Gaussian, we have 
\begin{equation}
	\alpha^3= \frac{\gamma^4}{\sqrt{\pi}v^3\Gamma_m}. \label{alfacubo}
\end{equation} 

\begin{figure}[tb]
	\centering
	\includegraphics[width=\textwidth]{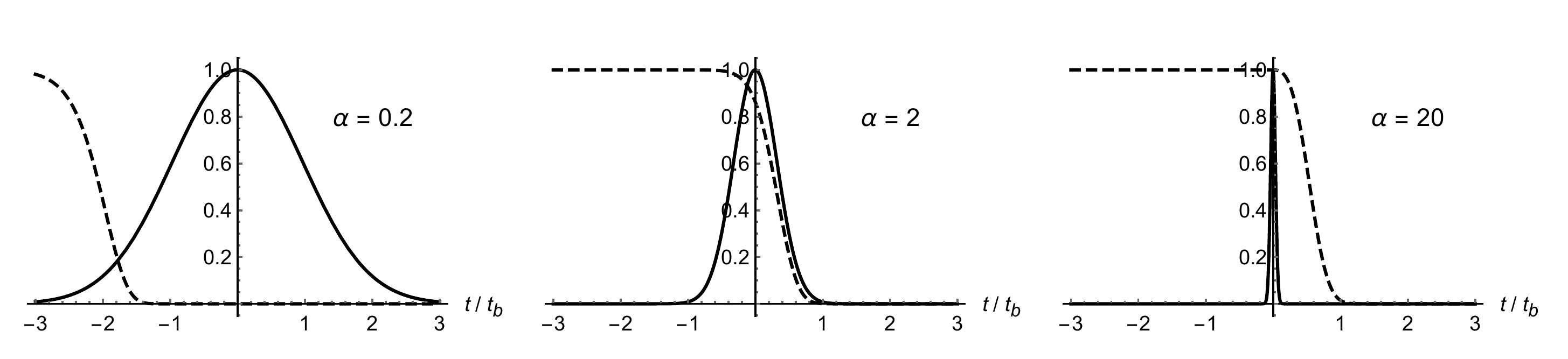}
	\caption{The nucleation rate $ \Gamma /\Gamma_m $ (solid lines) and the fraction of volume remaining in the high-temperature phase $ f_+ $ (dashed lines) for $ v=1 $.
		\label{figevol}}
\end{figure}

In Fig.~\ref{figevol} we consider the evolution of the phase transition
for different values of $ \alpha $. 
We see that, for $ \alpha\gtrsim 1 $,
the development of the phase transition takes place around the time $ t_m $ or later. 
Hence, in these cases the number density of nucleated bubbles is maximal (i.e., it is given by  the integral of the Gaussian),
and we have $ d_b\simeq d_{\min} =n_{\max} ^{-1/3}$ and the duration of the phase transition is given by 
$ t_b\simeq t_{\min} $.
Besides, we see that 
the parameter $ t_\Gamma $ gives an estimate of the duration of bubble nucleation.
In the limit $ \alpha\to\infty $, the time $ t_\Gamma $ becomes infinitely smaller than $ t_b $, and the nucleation rate becomes a delta function. 

In contrast, 
for small $ \alpha $ (left panel in Fig.~\ref{figevol}) the phase transition is completed before $ t=t_m $, and neither $ t_{\min} $ nor $ t_\Gamma $ give a correct estimate for its duration.
The limit $ \alpha\to 0 $ corresponds either  to $ \gamma\to 0 $ or to $ \Gamma_m \to\infty $.
Since the phase transition takes place away from the maximum of the Gaussian, we expect that 
the usual exponential approximation should be valid in this limit. 
To investigate this, we notice that, for any $ t_* $, we may write 
$ t-t_m=t-t_*+t_*-t_m $ in Eq.~(\ref{gammagaussnormal}), and we obtain
\begin{equation}
	\Gamma(t)=\Gamma_{*}e^{\beta(t-t_*)-\gamma^{2}(t-t_*)^{2}}, \label{Gammadesplazado}
\end{equation} 
where 
\begin{equation}
	 \Gamma_*=\Gamma_m e^{-\gamma^2(t_m-t_*)^2} ,\quad  \beta=2\gamma^2(t_m-t_*) 
\end{equation}
(notice that $ \beta $ depends on the time $ t_* $).
Let us consider  the time $ t_e $ at which $ I=1 $ ($ f_+= e^{-1}$).
The corresponding dimensionless variable $ x_e=(t_e-t_m) /t_\Gamma$ is given by the equation $ I(x_e,1) =\alpha^3$,
where $ I(x,\alpha) $ is given by Eq.~(\ref{Ix}).
For $ t_*=t_e $ we have $ \beta=2\gamma |x_e| $ and $ \Gamma_*=\Gamma_m e^{-x_e^2} $.
The function $ I (x,1)$ grows monotonically from $ I(-\infty,1)=0 $.
This can be seen more clearly from the definition of $ I (t)$, Eq.~(\ref{It}). 
Hence, for $ \alpha\to 0 $ we have $ x_e\to-\infty $.
This implies, in the first place,  that $ t_e-t_m \to-\infty$,  and, in the second place, that $ \gamma/\beta\propto|x_e|^{-1}\to 0 $,
so Eq.~(\ref{Gammadesplazado}) becomes indeed an exponential rate in this limit.
For large and negative $ x $ we have $ I(x,1)\simeq \sqrt{\pi}e^{-x^2}/2x^4$, so the equation for $ x_e $ becomes
$ \sqrt{\pi}e^{-x_e^2}/2x_e^4\simeq \alpha^3$, and using the relation (\ref{alfacubo}), we obtain
$ \Gamma_*\simeq 2\gamma^4 x_e^4/\pi v^3 =\beta^4/8\pi v^3$. Hence, in this limit Eq.~(\ref{Gammadesplazado})
coincides with our parametrization (\ref{nuclexpadim}),
\begin{equation}
	\Gamma(t)=\frac{\beta^{4}}{8\pi v^{3}}e^{\beta(t-t_{e})}.
	\label{nuclexpe}
\end{equation}

In Ref.~\cite{jlst17}, the GW spectrum was calculated for a nucleation
rate of the form  
\begin{equation}
	\Gamma(t)=H_{*}^4e^{\beta (t-t_*)-\gamma^{2}(t-t_*)^{2}}.
\end{equation} 
As we discussed in Sec.~\ref{gwwalls}, this model is motivated by an expansion of the instanton action $ S $ in powers of $ t-t_* $.
For many physical models we have 
$ \gamma\ll\beta $. 
Indeed, the usual approximation is to neglect $ \gamma $, which leads to the exponential nucleation rate. 
Notice that, in this parametrization, $ t_* $ is the time at which $ \Gamma=H_* ^4$, while the phase transition takes place at a later time, which is roughly given by $ \Gamma\sim \beta^4 $
(in general, $ \beta $ is a few orders of magnitude higher than $ H_* $).
Therefore, a new parametrization is used in \cite{jlst17},
\begin{equation}
	\Gamma(t)=\beta^{\prime 4} e^{\beta' (t-t_*')-\gamma^{2}(t-t_*')^{2}}.
	\label{gammajlst17}
\end{equation} 
Here, $ t_*' $ is the time at which $ \Gamma =\beta^{\prime 4}$, and 
for $ \gamma=0 $ we have $ \beta' =\beta$. 
On the other hand, for $ \gamma\neq 0 $ the nucleation rate (\ref{gammajlst17}) is a Gaussian and can be written in the form (\ref{gammagaussnormal}).
The parametrizations (\ref{gammajlst17}) and (\ref{gammagaussnormal}) are related by 
\cite{jlst17} 
\begin{equation}
	\Gamma_m = \beta^{\prime 4}e^{\beta^{\prime 2}/4\gamma^2}, \quad
	t_*'=t_m-\beta' /2\gamma^2.
	\label{tastpri}
\end{equation}
The results of Ref.~\cite{jlst17} depend on the ratio $ \gamma/\beta' $ and the velocity $ v $.
The relation with our dimensionless parameter $\alpha = \gamma t_b$  
is
\begin{equation}
	\sqrt{\pi}v^3(\gamma t_b)^3=(\gamma/\beta')^4 e^{-(\beta'/\gamma)^2/4}.
\end{equation}

We remark that, for $ \gamma\ll\beta $, the phase transition completes in a time of order $ \beta^{-1} $ 
well before the time $ t_m $ is reached.
Therefore, in the relevant time interval we may expand the exponential $ e^{-\gamma^{2}(t-t_*)^{2}} $
and obtain a perturbative expansion in powers of $ \gamma/\beta $, where each term of the expansion is computed using the exponential rate.
Thus, we obtain the lowest correction to the exponential case by writing Eq.~(\ref{Gammadesplazado}) as
\begin{equation}
	\Gamma(t)=\Gamma_* e^{\beta (t-t_*)}[1-\gamma^{2}(t-t_*)^{2}].
	\label{Gammaconcorreccion}
\end{equation} 
Taking $ t_*=t_e $, where now $ t_e $ is the time corresponding to $ f_+=1 $ for the {exponential rate}, 
the first factors in (\ref{Gammaconcorreccion}) are of the form (\ref{nuclexpe}).
The dimensionless nucleation rate is in this case   
\begin{equation}
	\tilde{\Gamma}(\tau)=(e^{\tau}/8\pi)[1-(\gamma/\beta)^2\tau^2]
\end{equation} 
(with $ \tau=\beta t $).
Hence, the correction to the exponential case is given by Eqs.~(\ref{Dsvconst})-(\ref{G}), with the polynomials $ P_i $  replaced by
$-(\gamma/\beta)^2 \tau_N^2 P_i $ in Eq.~(\ref{F}), and the same for $ Q_\pm $ in Eq.~(\ref{G}).
The integrals in these equations can be done analytically, and the functions $ F_i $ and $ G_\pm $ are of the form (\ref{FGfinal}),
where $ \tilde F_i $ and $ \tilde G $ are polynomials which are now more cumbersome than Eqs.~(\ref{Ftil0})-(\ref{Gtil}). 
In any case, computing this correction to the exponential case may be more useful than considering a Gaussian rate.
We shall investigate this kind of approximation elsewhere.

In the opposite case, in which the phase transition occurs around the time $ t_m $,
the parametrization (\ref{gammajlst17}) in terms of $ \beta' $ is no longer useful. 
In particular, 
according to Eqs.~(\ref{tastpri}), we have $ t_*'\to t_m $ only for $ \beta'\to 0 $ or $ \gamma\to\infty $.
In practice,
we may have $ \beta'>\gamma $ and the bubble nucleation still occur in a time $ \gamma^{-1} $ around $ t_m $, while 
the reference time $ t_*' $ may fall outside the relevant range.
Therefore, if the value of $ \gamma $ is calculated by expanding $ S $ around the corresponding temperature $ T_*' $, 
the error may be large.
Let us consider some specific examples from physically motivated models. 
In the case of strong supercooling considered in Ref.~\cite{mr17}
we have typical values $ \gamma/H_*\sim 10,\Gamma_m/H_*^4\sim 1000$. 
This gives, according to Eq.~(\ref{tastpri}), a ratio $ \beta'/\gamma $ in the range $ 2-3 $ and 
$ t_m-t_*'\simeq \gamma^{-1} $. 
On the other hand, in the case of reheating, for the physical model considered in Ref.~\cite{mr18}
we have $ \Gamma_m/H_*^4 \sim 10^{17}-10^{18}$ and $ \gamma/H_*\sim 10^3-10^4 $. 
This gives $ \beta'/\gamma\simeq 9-10 $ and $ t_m-t_*'\simeq 5\gamma^{-1} $.

\section{The shape of the spectrum}

\label{shapes}

The GW spectrum for both the exponential nucleation and the simultaneous
nucleation were computed in Ref.~\cite{w16} for $v=1$, and Fig.~3
of that work can be directly compared with our solid and dashed curves
for $v=1$ in Fig.~\ref{figspectra}. The results are in qualitative
agreement for the shapes of the curves as well as for the relative
positions of the peak for the two models. Quantitatively, though,
our results have order 1 differences with those simulations. Nevertheless,
for the exponential case, our results are in agreement with more recent
simulations \cite{k18} as well as with the semi-analytic treatment
of Ref.~\cite{jt17}.

\begin{figure}[tbh]
\centering 
\includegraphics[width=0.5\textwidth]{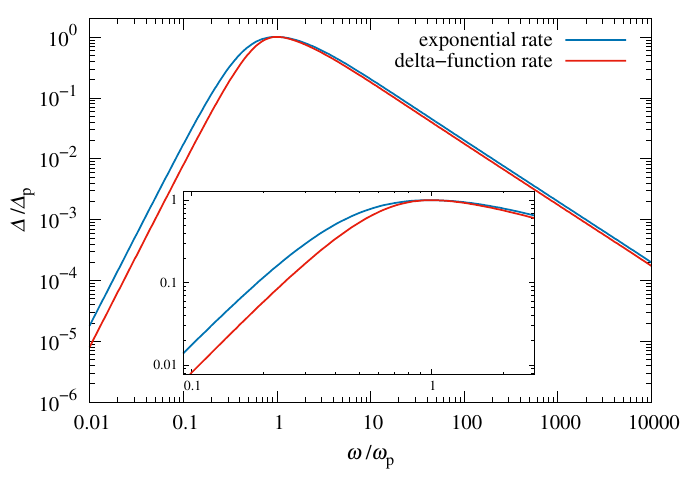}%
\includegraphics[width=0.5\textwidth]{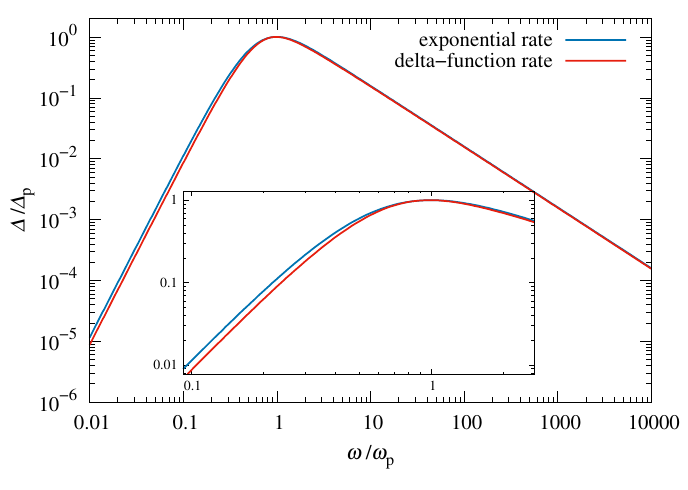}
\caption{The spectral shape for the exponential and the delta-function rates,
for $v=1$ (left) and $v=0.3$ (right). \label{figjapos}}
\end{figure}

The numerical results of Ref.~\cite{jlst17} for the Gaussian case were presented with 
the frequency and amplitude of the GW spectrum $ \Delta $ 
normalized with their peak values $ \omega_p,\Delta_p $.
In this way, the maximum of the spectra for different model parameters coincide.
The information on the values of $ \omega_p $ and $ \Delta_p $ is lost, but
the shape of the spectra can be directly compared.
We consider a similar plot in Fig.~\ref{figjapos} for the two limiting cases of the 
Gaussian nucleation rate, namely, $ \alpha\to 0 $ (exponential rate)
and $ \alpha\to\infty $ (delta-function rate).
The curves for different values of $ \alpha $ fall between these two cases.
Only the range $0.1\lesssim\omega/\omega_{\mathrm{p}}\lesssim3$
was considered in Ref.~\cite{jlst17}, due to numerical difficulties
of their multi-dimensional integration. The inset in Fig.~\ref{figjapos}
shows this range for a better comparison. It can be appreciated that
these curves are in agreement with those of \cite{jlst17}. 

\bibliographystyle{jhep}
\bibliography{papers}

\end{document}